\documentclass[useAMS,usenatbib,usegraphicx]{mn2e}

\usepackage{aas_macros}

\usepackage{fixltx2e}
\usepackage{multirow}
\usepackage{color}

\title[ALMA and the dynamics of protostellar discs]{Revealing the dynamics of Class 0 protostellar discs with ALMA}
  \author[D. Seifried et al.]
  {D.~Seifried,$^{1,2}$\thanks{seifried@ph1.uni-koeln.de}, \'A.~S\'anchez-Monge,$^{1}$ S. Walch,$^{1}$ R.~Banerjee$^{2}$ \\
  $^1$I. Physikalisches Institut, Universit\"at zu K\"oln, Z\"ulpicher Str. 77, 50937 K\"oln, Germany\\
  $^2$Hamburger Sternwarte, Universit\"at Hamburg, Gojenbergsweg 112, 21029 Hamburg, Germany\\}

\date{Released 2016}

\pagerange{\pageref{firstpage}--\pageref{lastpage}} \pubyear{2015}

\begin{document}

\label{firstpage}

\maketitle

\begin{abstract}
We present synthetic ALMA observations of Keplerian, protostellar discs in the Class 0 stage studying the emission of molecular tracers like $^{13}$CO, C$^{18}$O, HCO$^+$, H$^{13}$CO$^+$, N$_2$H$^+$, and H$_2$CO. We model the emission of discs around low- and intermediate-mass protostars. We show that under optimal observing conditions ALMA is able to detect the discs already in the earliest stage of protostellar evolution, although the emission is often concentrated to the innermost 50 AU. Therefore, a resolution of a few 0.1'' might be too low to detect Keplerian discs around Class 0 objects. We also demonstrate that under optimal conditions for edge-on discs Keplerian rotation signatures are recognisable, from which protostellar masses can be inferred. For this we here introduce a new approach, which allows us to determine protostellar masses with higher fidelity than before.
Furthermore, we show that it is possible to reveal Keplerian rotation even for strongly inclined discs and that ALMA should be able to detect possible signs of fragmentation in face-on discs. In order to give some guidance for future ALMA observations, we investigate the influence of varying observing conditions and source distances. We show that it is possible to probe Keplerian rotation in inclined discs with an observing time of 2 h and a resolution of 0.1'', even in the case of moderate weather conditions. Furthermore, we demonstrate that under optimal conditions, Keplerian discs around intermediate-mass protostars should be detectable up to kpc-distances.
\end{abstract}


\section{Introduction}

In the past decade the formation of protostellar discs in the earliest, embedded stage of protostellar evolution, the so-called Class 0 stage \citep{Andre93}, has been discussed extensively in literature from both the observational and numerical modelling side. From observations of the more evolved Class I/II stage protostars it is known that the fraction of discs in these later stages is well above 50 percent \citep[see e.g. the review by][]{Williams11}, which hints towards the existence of discs right from the onset of protostar formation. Furthermore, the detection of protostellar outflows strongly indicates the presence of discs in the earliest stage of protostellar evolution \citep{Arce07}. For this reason, an increasing effort has been made over the last years to directly identify the presence and dynamics of protostellar discs around Class 0 objects, which resulted in the detection of several discs around low- and intermediate-mass protostars, which are possibly in Keplerian rotating \citep{Tobin12,Hara13,Murillo13,Sanchez13,Yen13,Codella14,Ohashi14}. This is in agreement with previous observations based on SED modelling, which suggest the presence of well-defined, rotationally supported discs \citep{Jorgensen09,Enoch11}. In contrast, \citet{Maury10}, \citet{Maret14} as well as \citet{Yen13} (for a part of their observed objects) do not find Keplerian discs around Class 0 objects. This raises the question what the actual fraction of (rotationally supported) discs in the Class 0 stage is. A fraction well below 1, i.e. not all young protostars have a protostellar disc, would imply a successive growth of centrifugally supported discs during the evolution towards the Class I/II phase \citep[e.g.][]{Dapp10,Dapp12}.

Alternatively, if typical disc sizes are not larger than \mbox{$\sim$ 100 AU} in diameter, the non-detection of discs could be a resolution issue. Indeed, recent simulations of protostellar disc formation including turbulence and magnetic fields show that discs usually do not extend over more than about 100 AU in size \citep{Myers13,Joos13,Seifried12,Seifried13,Seifried15}. Similar results were found by \citet{Walch10,Walch12} showing that only in the presence of large-scale turbulent motions discs with radii larger than 10 AU develop around low-mass protostars. Furthermore, discs around Class II protostellar sources are in general also found to be not significantly larger than 100 AU \citep[e.g.][]{Williams11}. Assuming that discs grow during their lifetime due to viscous spreading \citep[e.g.][]{Nakamoto94}, this puts some constraints on the disc size in the Class 0 stage. Hence, both numerical simulations and observations suggest that in typical protostellar cores including turbulent motions, disc sizes should be of the order of 10 -- 100 AU, which -- at source distances of a few 100 pc -- would require a resolution better than $\sim$ 0.1''. With ALMA such highly resolved observations, which can reveal the dynamics of young protostellar discs in great detail, are now accessible.

In this work we present synthetic ALMA observations of self-consistent simulations of protostellar discs for various molecular line transitions and directly compare them with the simulation data. This allows us to study to what extent discs and Keplerian rotation profiles can be detected around low- and intermediate-mass Class 0 protostars and, if detected, how reliable the inferred disc parameters are. Moreover, we will investigate which observational conditions like observing time, weather conditions, angular resolution, and molecular tracers are required to study protostellar discs with ALMA, in particular for the upcoming cycles, where the full capacity of ALMA will be available.

Main outcomes of this work are that with ALMA it will be possible to identify Keplerian rotation for most of the lines considered, when adopting an observing time of 2 -- 5 h per object, a resolution of 0.02 -- 0.1'', and weather conditions corresponding to a precipitable water vapour of 0.5 -- 1 mm. Moreover, we show that protostellar masses inferred from PV diagrams are in general accurate within \mbox{$\sim$ 10 percent.}

The structure of this paper is as follows: In Section~\ref{sec:IC} we briefly describe the 3D-magnetohydrodynamical simulations and their main outcomes. The radiative transfer simulations as well as the generation of the synthetic ALMA observations are described in Section~\ref{sec:synobs}. In Section~\ref{sec:results} we discuss the synthetic data obtained by assuming ideal observing conditions. We study whether Keplerian rotation signatures can be identified, how reliable protostellar masses inferred from position-velocity diagrams are, and how the inclination angle influences the results. In Section~\ref{sec:conditions} we investigate the influence of varying observing conditions, and give a guideline for future observations. The results are discussed in Section~\ref{sec:discussion}, before we conclude in Section~\ref{sec:conclusions}.

\section{Overview of the simulations}
\label{sec:IC}

We carried out magnetohydrodynamical (MHD) simulations of the collapse of magnetized, turbulent, prestellar cores to follow the formation of protostellar discs. For this we use the adaptive-mesh refinement code FLASH \citep{Fryxell00} in version 2 using a MHD solver developed by \citet{Bouchut07} and \citet{Waagan11} with a maximum spatial resolution of 1.2 AU. The Poisson equation for gravity is solved with a Barnes-Hut tree code \citep{Wunsch15}. In addition, we make use of the sink particle routine \citep{Federrath10} to model the formation of protostars. The simulations were discussed in detail in \citet{Seifried12,Seifried13,Seifried15}, which is why here we only briefly describe the basic setup of the simulations. For more details we refer the reader in particular to \citet{Seifried13}.

In this paper we consider two simulations, one with an initial core mass of 100 M$_{\sun}$ and one with 2.6 M$_{\sun}$. Throughout this work we refer to them as model 1 and 2, respectively. The cores have a radius of 0.125 pc (model 1) and 0.0485 pc (model 2). The density profile in model 1 declines as
\begin{equation}
 \rho \propto r^{-1.5} \, ,
\end{equation}
levelling off at 0.0125 pc, whereas that of model 2 corresponds to that of a Bonnor-Ebert sphere. The initial temperature of the gas is 20 K and 15 K for model 1 and 2, respectively, reflecting the somewhat lower temperatures typically observed in low-mass cores \citep[e.g.][]{Ragan12,Launhardt13,Sanchez13b}. We use a cooling function, which includes the effects of molecular cooling and dust cooling as well as a treatment for optically thick gas \citep[see][for details]{Banerjee06b}. The cores are threaded by a magnetic field parallel to the $z$-axis with the strength chosen such that the (normalised) mass-to-flux ratio $\mu$ of the cores is 2.6. We emphasise that $\mu$ is given in units of the critical mass-to-flux ratio $\mu_\rmn{crit} = 0.13/\sqrt{G}$ \citep{Mouschovias76}, where $G$ is the gravitational constant.

In all cores a turbulent velocity field is present with the turbulent Mach number of 2.5 for the intermediate-mass model 1 and 0.74 for the low-mass model 2 \citep[e.g.][]{Sanchez13b}. Finally we note that for neither of the two models considered here an initial coherent rotation is present. We list all important parameters in Table~\ref{tab:models}.
\begin{table}
  \caption{Overview of the initial parameters of model 1 and 2.}
 \label{tab:models}
 \centering
 \begin{tabular}{lcc}
 \hline
   & model 1 & model 2 \\
 \hline
 core mass [M$_{\sun}$] & 100 & 2.6 \\
 core radius [pc] & 0.125 & 0.485 \\
 temperature [K] & 20 & 15 \\
 Mach number & 2.5 & 0.74 \\
 mass-to-flux-ratio & 2.6 & 2.6 \\
 m$_\rmn{sinks}$ [M$_{\sun}$] & 15.3 & 0.62 \\
 \hline
 \end{tabular}
\end{table}

\begin{figure}
\centering
\includegraphics[width=0.49\linewidth]{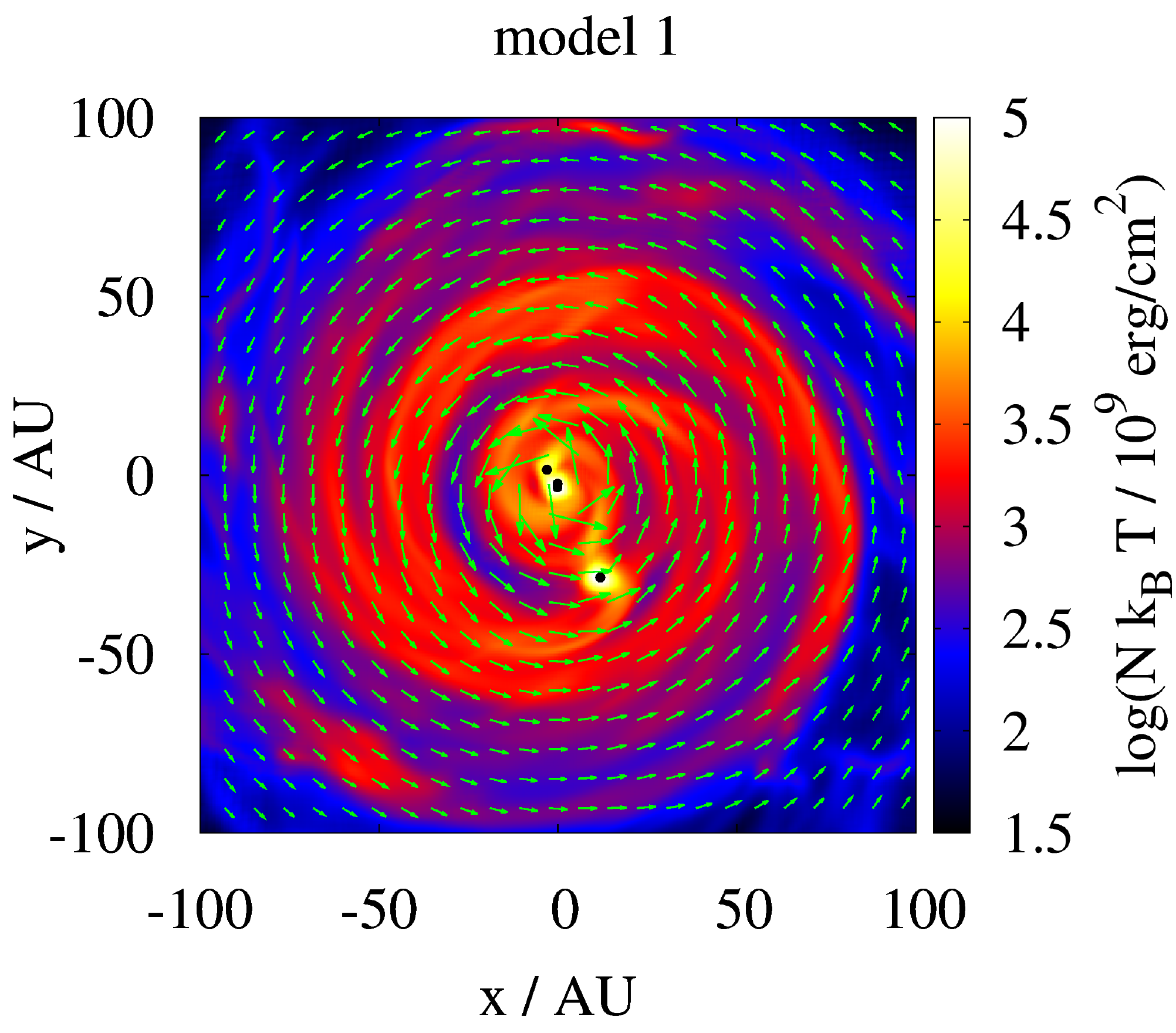}
\includegraphics[width=0.49\linewidth]{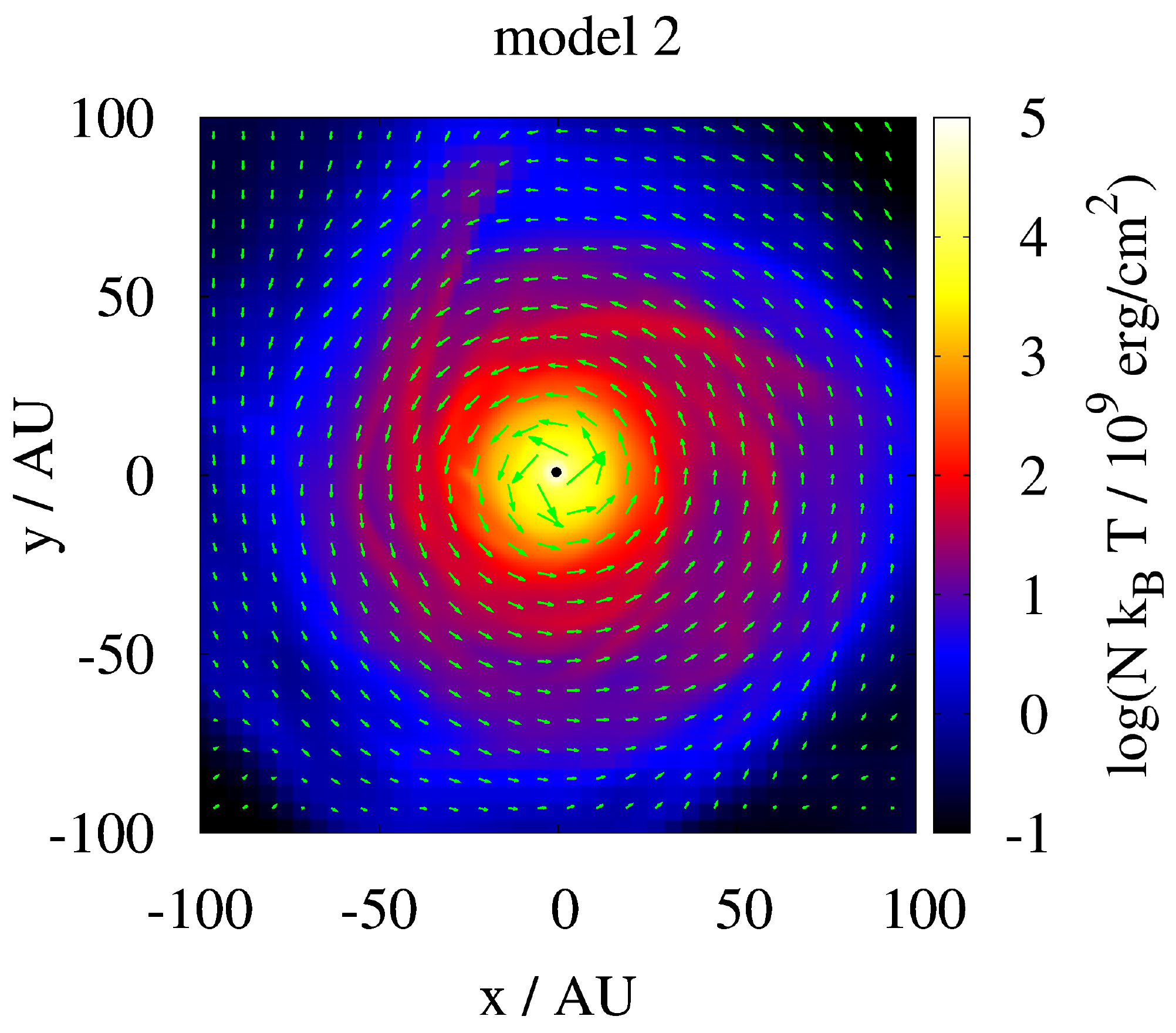} \vspace{0.1cm}\\
\includegraphics[width=0.49\linewidth]{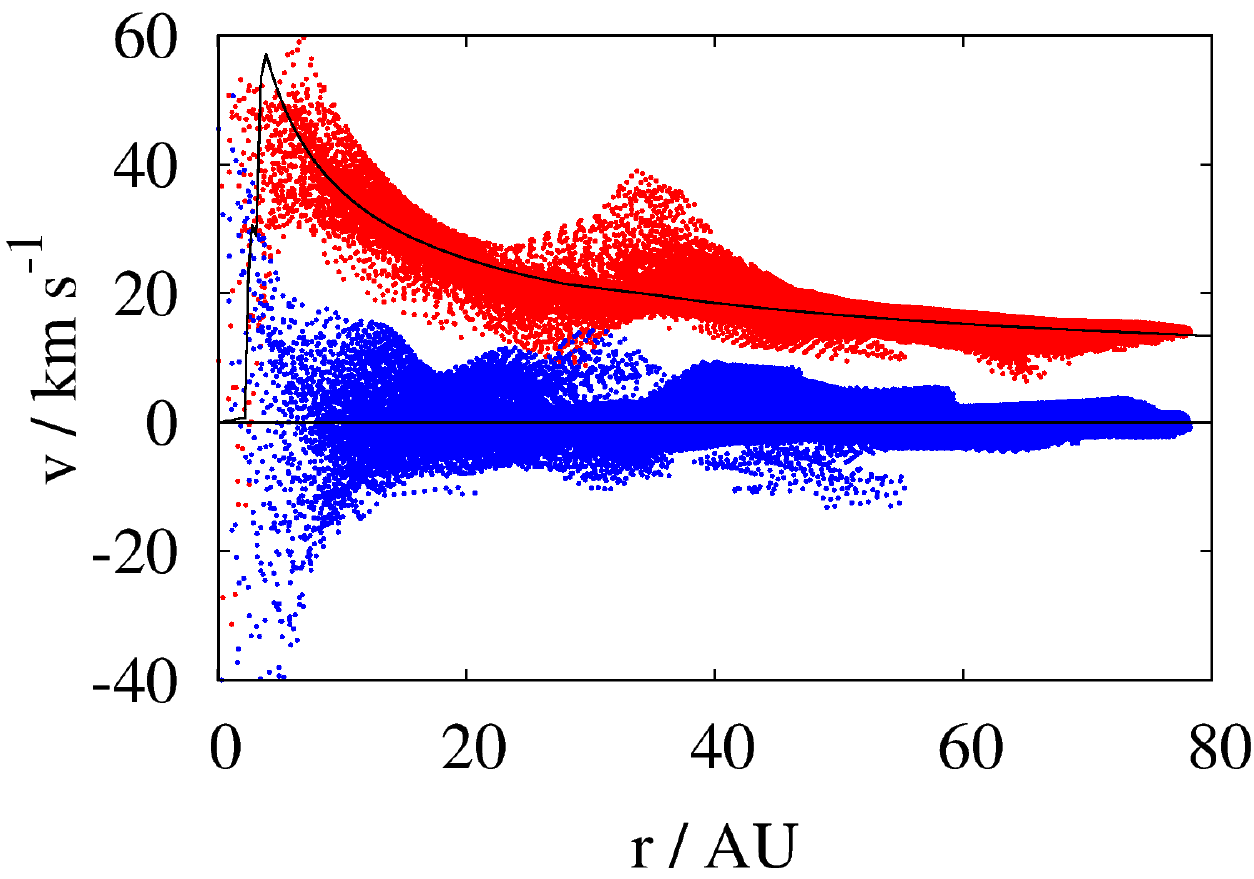}
\includegraphics[width=0.49\linewidth]{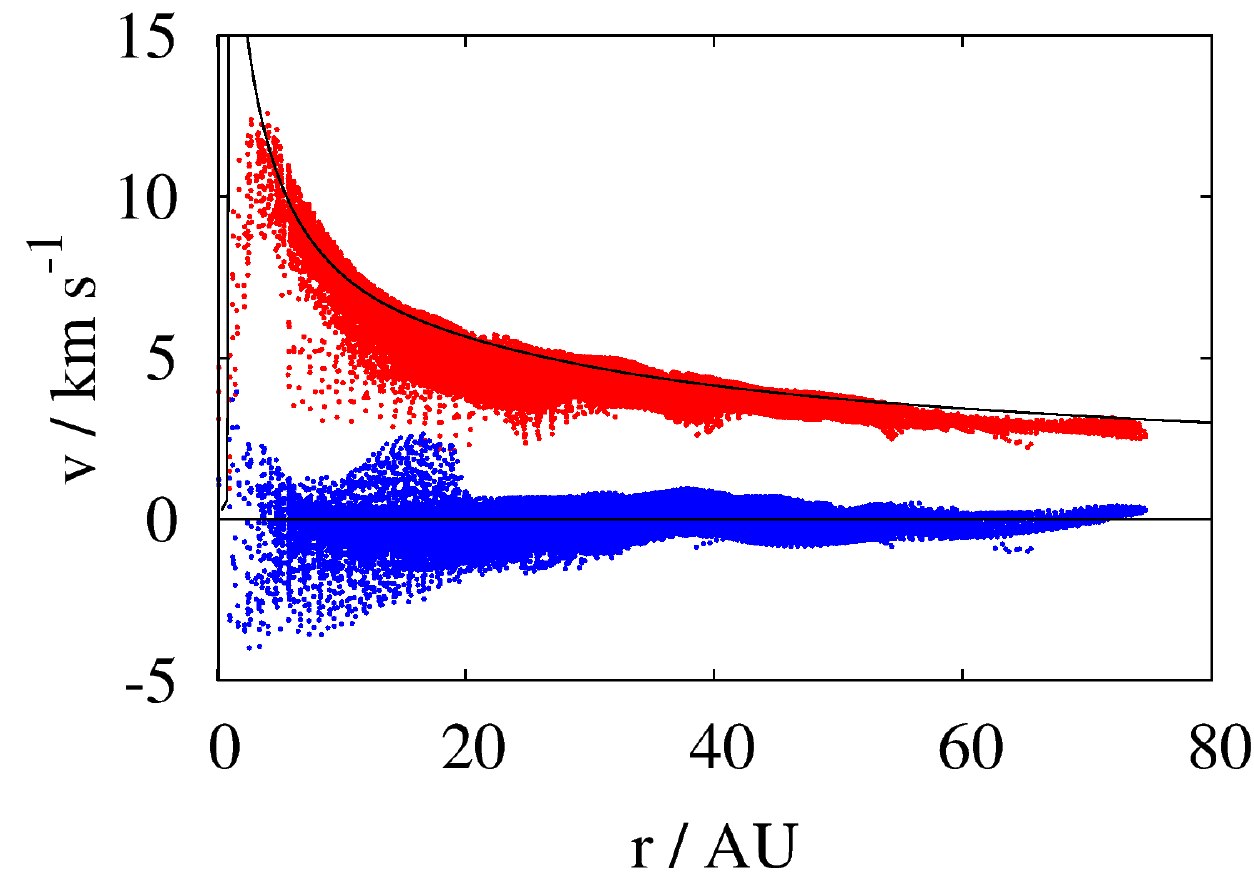}
\caption{Top: Surface density of the thermal energy Nk$_B$T of the disc formed in model 1 at $t$ = 25 kyr after the first sink particle has formed (left) and of the disc in model 2 at $t$ = 30 kyr after the formation of the first sink particle. In addition we show the line-of-sight averaged velocity field (green arrows) and the protostars (black dots). Bottom left panel: Radial dependence of the rotation velocity (red), the radial velocity (blue), and the Keplerian velocity (upper black line).}
\label{fig:simulation}
\end{figure}

Next, we briefly recapitulate the main outcomes of the two simulations presented in this work \citep[again, see][for more details]{Seifried12,Seifried13,Seifried15}. In both runs Keplerian discs are formed within a few kyr after the formation of the central protostar. 
In Fig.~\ref{fig:simulation} we show the discs 25 kyr (model 1) and 30 kyr (model 2) after the first protostar has formed, which also corresponds to the times at which the synthetic observations are produced. We emphasise that in the top row we plot the surface density of the thermal energy Nk$_B$T, which -- at a very basic level -- should resemble the emission emitted from the disc.
From the profile of the thermal energy, it seems problematic to define a disc size since the profile falls off relatively smoothly towards larger radii. Therefore, we can only roughly estimate the disc mass and diameter, which range from about 50 -- 150 AU and 0.05 up to a few 0.1 M$_{\sun}$, respectively.
However, it can be clearly seen, that up to radii $\geq$ 70 AU the discs stay Keplerian and the infall velocity is smaller than the rotation velocity (bottom panel of Fig.~\ref{fig:simulation}). The disc in model 1 contains 4 protostars with masses ranging from 0.9 to 9.3 M$_{\sun}$ and a total mass in stars of 15.3 M$_{\sun}$. The disc in model 2 has not fragmented until we stop the simulation and contains a single low-mass protostar with a mass of 0.62 M$_{\sun}$. 

Furthermore, we find that in the initial phase of disc formation the mass-to-flux ratio stays well below a value of 10. Therefore, it is below the critical value for which Keplerian disc formation is suppressed in non-turbulent simulations \citep{Allen03,Matsumoto04,Machida05,Banerjee06,Banerjee07,Price07,Hennebelle08,Hennebelle09,Duffin09,Commercon10,Burzle11,Seifried11}, which is why flux-loss alone cannot explain the formation of Keplerian discs for our turbulent runs. In \citet{Seifried12,Seifried13,Seifried15} we argue that the formation of rotationally supported discs at such early phases is rather due to the disordered magnetic field structure and due to the highly anisotropic accretion structure with accretion mainly occurring through a few very narrow channels. We argue that these effects lower the classical magnetic braking efficiency with respect to a highly idealized, rotating, non-turbulent system with an ordered magnetic field of comparable strength.

\section{Synthetic observations}
\label{sec:synobs}

We use the freely available radiative transfer code RADMC-3D \citep{Dullemond12} to produce synthetic line emission maps of different molecular transitions for both models. For model 1 we take a snapshot at $t$ = 25 kyr after the first sink particle has formed, for model 2 a snapshot at $t$ = 30 kyr after the initial sink formation. This corresponds to the situation displayed in Fig.~\ref{fig:simulation} (model 1 in the left panel, right: model 2). 

In this paper we consider various transitions of the molecules $^{13}$CO, C$^{18}$O, HCO$^+$, H$^{13}$CO$^+$, N$_2$H$^+$, as well as ortho- and para-H$_2$CO. For all molecules we assume a fixed abundance relative to molecular hydrogen. For $^{13}$CO and C$^{18}$O we used relative abundances of 1.3 $\times$ 10$^{-6}$ and 1.8 $\times$ 10$^{-7}$, respectively \citep{Wilson94}. For N$_2$H$^+$ we have adopted a fiducial value of 3 $\times$ 10$^{-10}$ \citep{Caselli02,Jorgensen04,Tafalla04,Beuther09,Johnstone10,Morales12,Tsitali15} and for HCO$^+$ a value of 1 $\times$ 10$^{-9}$ \citep{Jorgensen04,Sandell10,Morales12,Zernickel12}. For the relative abundance of H$^{13}$CO$^+$ we have taken the abundance of HCO$^+$ scaled down by a factor of 77 typical for the ratio of $^{12}$C to $^{13}$C in the local interstellar medium \citep{Wilson94}. For H$_2$CO we assume that its abundance is increased towards the disc due to evaporation from icy dust grains. Within a radius of 100 AU from the disc we use an abundance of 2 $\times$ 10$^{-7}$, whereas in the outer parts we use an abundance of 2 $\times$ 10$^{-10}$, which is in rough agreement with observations \citep[e.g.][]{Ceccarelli04,Maret04}. All lines considered in this work are  listed in Table~\ref{tab:lines}.
\begin{table*}
  \caption{Overview of the produced line emission maps for the different molecules and transitions. The numbers give the viewing angles for which the maps were calculated.}
 \label{tab:lines}
 \begin{tabular}{cccccc}
 \hline
 line, molecule & $^{13}$CO 						& C$^{18}$O 						& HCO$^+$ 							& H$^{13}$CO$^+$ 						& N$_2$H$^+$ \\
 \hline
  $J$ = 1--0 & 90$^{\circ}$, 60$^{\circ}$, 30$^{\circ}$, 0$^{\circ}$	& 90$^{\circ}$, 60$^{\circ}$, 30$^{\circ}$, 0$^{\circ}$	& 90$^{\circ}$, 60$^{\circ}$, 30$^{\circ}$, 0$^{\circ}$	& 90$^{\circ}$, 60$^{\circ}$, 30$^{\circ}$, 0$^{\circ}$	& 90$^{\circ}$, 60$^{\circ}$, 30$^{\circ}$, 0$^{\circ}$ \\
  $J$ = 2--1 & 90$^{\circ}$, 60$^{\circ}$, 30$^{\circ}$, 0$^{\circ}$	& 90$^{\circ}$, 60$^{\circ}$, 30$^{\circ}$, 0$^{\circ}$	& --- & ---& --- \\
  $J$ = 3--2 & 90$^{\circ}$						& 90$^{\circ}$						& 90$^{\circ}$, 60$^{\circ}$, 30$^{\circ}$, 0$^{\circ}$	& 90$^{\circ}$, 60$^{\circ}$, 30$^{\circ}$, 0$^{\circ}$	& 90$^{\circ}$, 60$^{\circ}$, 30$^{\circ}$, 0$^{\circ}$ \\
  $J$ = 4--3 & ---							& ---							& 90$^{\circ}$						& 90$^{\circ}$, (model 1: 60$^{\circ}$, 30$^{\circ}$, 0$^{\circ}$)						& 90$^{\circ}$ \\
  \hline
  \hline
  molecule ($\nu$) & o-H$_2$CO (225.7 GHz) & p-H$_2$CO (145.6 GHz) & p-H$_2$CO (218.2 GHz) & p-H$_2$CO (362.7 GHz) \\
    \hline
  & 90$^{\circ}$ & 90$^{\circ}$ & 90$^{\circ}$ & 90$^{\circ}$ \\
  \hline
 \end{tabular}
\end{table*}

The critical densities of the different transition lines given in Table~\ref{tab:lines} lie between $\sim$ 10$^{3}$ cm$^{-3}$ and $\sim$ 10$^{7}$ cm$^{-3}$. Given the fact that densities in protostellar discs as well as in their immediate surroundings are significantly higher than these densities, for the calculation of the line intensity we have assumed local thermodynamic equilibrium. The molecular data, e.g. the Einstein coefficients are take from the Leiden Atomic and Molecular database \citep{Schoier05}.

The emission maps produced with RADMC-3D cover a velocity range from --60 km s$^{-1}$ to 60 km s$^{-1}$ for model 1 and --20 km s$^{-1}$ to 20 km s$^{-1}$ for model 2, both centred at a systemic velocity of 0 km s$^{-1}$, which guarantees that all emission is captured properly (see bottom panels of Fig.~\ref{fig:simulation}). We use channel widths of 0.2 km s$^{-1}$, which results in 601 and 201 channels for model 1 and 2, respectively. This high spectral resolution allows us to properly track the rotational motions of the discs, but also leads to significant computational cost for the production of the emission maps. The standard viewing angle is 90$^{\circ}$, which corresponds to a disc seen edge-on. A viewing angle of 0$^{\circ}$, on the other hand, corresponds to a disc seen face-on. In addition, we also produce images for an inclination of 60$^{\circ}$ and 30$^{\circ}$ (see Table~\ref{tab:lines} for an overview over all models considered).

In a next step, the emission maps are post-processed to simulate the effects of a real observation with ALMA on the model images. The post-processing has been done using the CASA\footnote{CASA stands for Common Astronomy Software Applications and can be downloaded from http://casa.nrao.edu/} software \citep{McMullin07}, version~4.3.1. The first step is to produce the visibilities of the model image by using the task \texttt{simobserve}. We have considered a source distance of 150~pc for both models. In addition, for model 1 we also consider a larger distance of 1~kpc (see Section~\ref{sec:distance}). We produce visibility datasets using different observational conditions: (i) an observing time of 2 and 5~h, (ii) the highest possible angular resolution at the corresponding frequency (with a limit of 0.02'' for the highest frequencies) and with a coarser resolution of 0.1'', and (iii) different weather conditions determined by the amount of precipitable water vapor (pwv), which is set to either 0.5~mm or 1~mm. The sampling time was set to 10 seconds, and we assume the source is observed during its transit. Regarding the frequency setup, we consider the rest frequency of each molecular transition, we set the line-of-sight velocity of the source to be 0~km~s$^{-1}$, and consider channel widths of 0.2~km~s$^{-1}$. In a second step, the visibility datasets are cleaned with the task \texttt{clean} to produce the images used in the analysis. Channel maps are created for all the molecular transitions listed in Table~\ref{tab:lines}, with a natural weighting to improve the sensitivity and a gain factor of 0.05 to better clean extended emission. In Table~\ref{tab:casa} we list the typical synthesised beams as well as the rms noise levels per channel of the final images for the different frequency bands considered in this work: band 3 (3~mm), band 4 (2~mm), band 6 (1~mm), and band 7 (0.8~mm).
\begin{table}
 \caption{Overview of the observational parameters of the synthetic emission maps created with CASA.}
 \label{tab:casa}
 \begin{tabular}{ccc}
  \hline
  band & synthesised beam [''] & rms [mJy~beam$^{-1}$] \\
  \hline
  3 & 0.045 -- 0.060 & 1.2 \\
  4 & 0.035 -- 0.037 & 1.0 \\
  6 & 0.022 -- 0.024 & 1.1 \\
  7 & 0.020 -- 0.023 & 1.8 \\
  \hline
 \end{tabular}
\end{table}

\section{Results}
\label{sec:results}

We first present the integrated intensity maps of discs seen edge-on, i.e. under a viewing angle of 90$^{\circ}$, for a number of selected lines. In a second step, we infer the dynamics of these discs and analyse to what extent Keplerian rotation signatures can be observed for the different molecular tracers. We also examine what protostellar masses are obtained from the observed velocity structure. Finally, we investigate the influence of the viewing angle, in particular how this affects the inferred disc dynamics and whether ALMA is able to probe disc fragmentation. We note that the synthetic observations discussed in the following were produced assuming the best angular resolution for the corresponding frequency (ranging from 0.02'' -- 0.05'', see Table~\ref{tab:casa}), an observing time of 5 h, and a precipitable water vapour (pwv) of 0.5 mm, which corresponds to optimal observing conditions.

\subsection{Intensity and  first moment maps}
\label{sec:maps}

\begin{figure*}
\includegraphics[width=\linewidth]{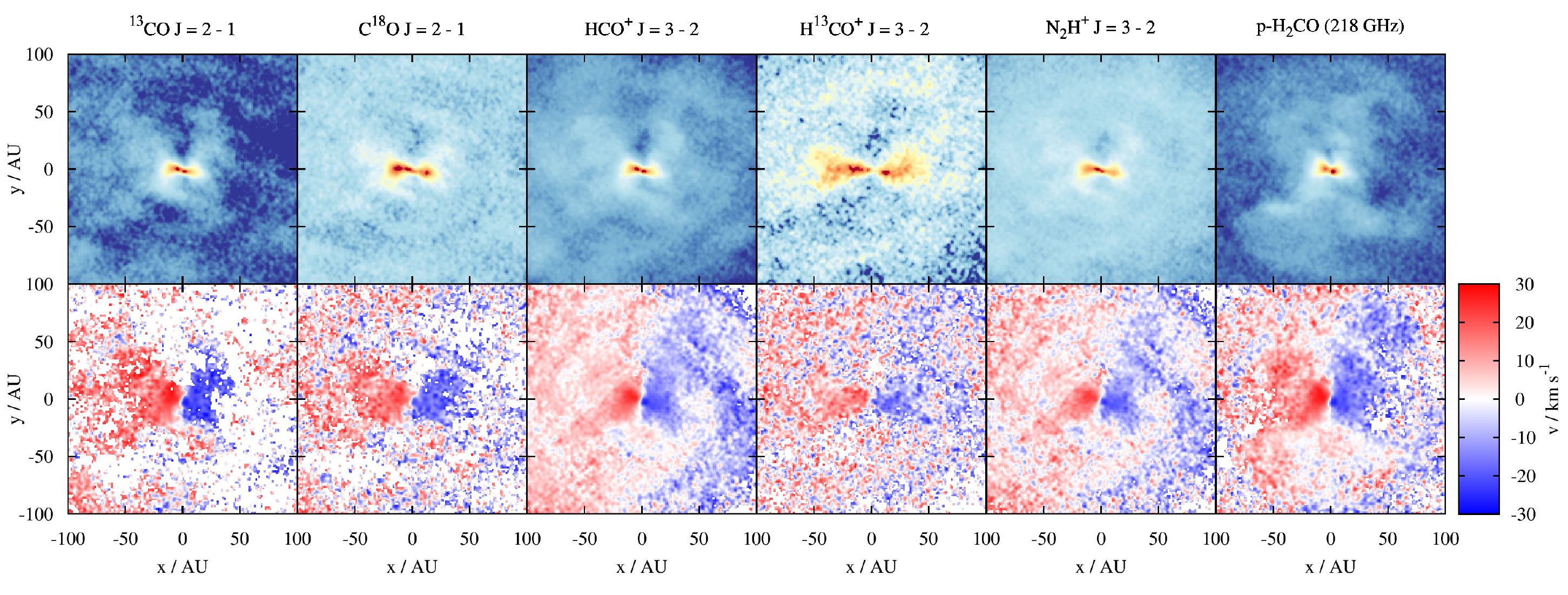}
\caption{Top Row: Integrated intensities (0th-order moment maps) of $^{13}$CO $J$ = 2--1, C$^{18}$O $J$ = 2--1, HCO$^+$ $J$ = 3--2, H$^{13}$CO$^+$ $J$ = 3--2, N$_2$H$^+$ $J$ = 3--2, and p-H$_2$CO (218 GHz) (from left to right) for model 1 (intermediate-mass star) obtained under the assumption of an optimal spatial resolution (see Table~\ref{tab:casa}) and a distance of 150 pc. Note that the intensities scales are different in each panel. Bottom row: same as top row but for the 1st-order moment map. A clear velocity gradient is observable in all cases.}
\label{fig:intensity}
\end{figure*}

\begin{figure*}
\includegraphics[width=\linewidth]{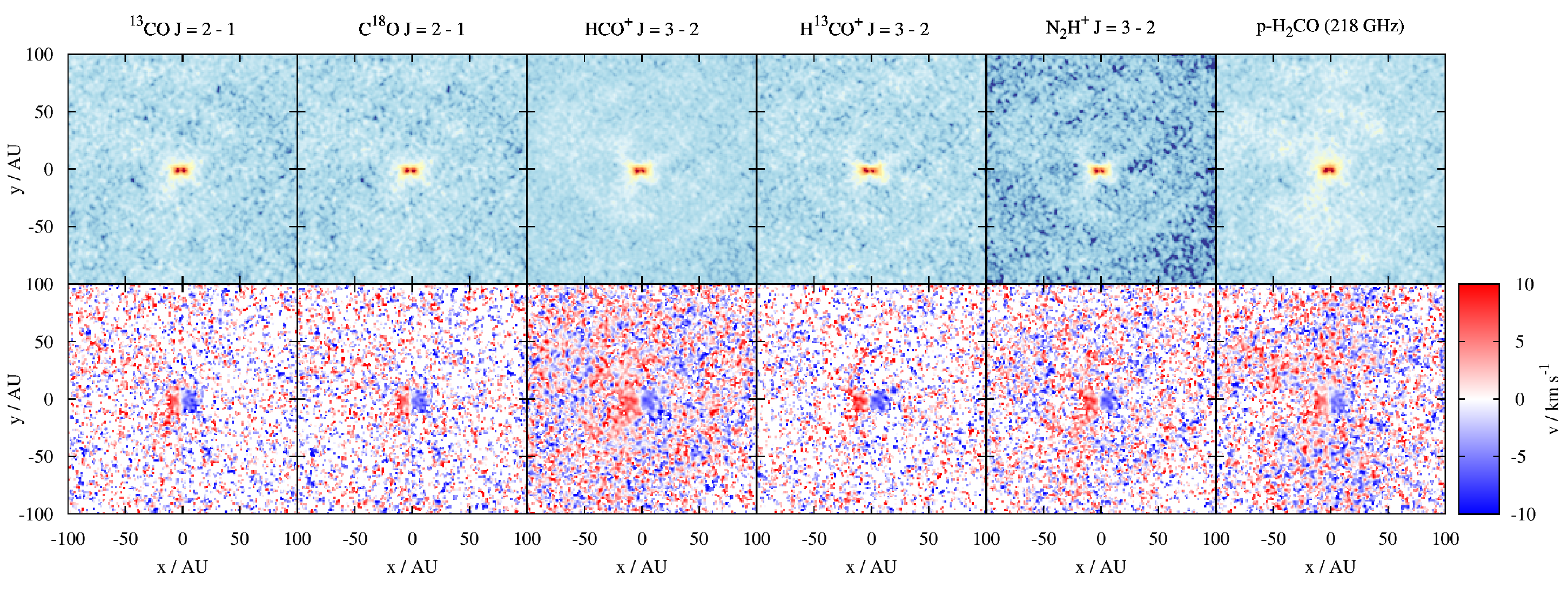}
\caption{Same as in Fig.~\ref{fig:intensity} but for model 2.}
\label{fig:intensity2}
\end{figure*}

First, we investigate the integrated intensity as well as the first moment maps, i.e. the velocity-integrated intensity maps, for $^{13}$CO $J$ = 2--1, C$^{18}$O $J$ = 2--1, HCO$^+$ $J$ = 3--2, H$^{13}$CO$^+$ $J$ = 3--2, N$_2$H$^+$ $J$ = 3--2, and p-H$_2$CO (218 GHz) in Fig.~\ref{fig:intensity} (model 1) and Fig.~\ref{fig:intensity2} (model 2). In both figures, the integrated emission (0th-order moment map) is shown in the top row, the 1st-order moment map in the bottom row. For both models, the emission is highly concentrated and is dominated by the channels close to the systemic velocity\footnote{The systemic velocity is set to zero in both models.}. Elongated disc structures are visible in both models and for all lines, with the bulge of the emission coming from a region smaller than 100 AU. In particular, for the low-mass protostellar model 2 the emission seems to extend not further than a few 10 AU. Hence, the observations generally probe a region \textit{smaller} than the part of the disc that is in Keplerian rotation ($\sim$ 70 AU, see Fig.~\ref{fig:simulation}). The strongest contrast between the central emission and the surrounding emission is obtained for $^{13}$CO. However, even for the least abundant molecule H$^{13}$CO$^+$ the disc is clearly visible.

The first moment maps of the considered lines (bottom rows of Figs.~\ref{fig:intensity} and~\ref{fig:intensity2}) show clear velocity gradients. These indicate rotational motions with velocities ranging up to $\sim$ 30 km s$^{-1}$ for model 1 and $\sim$ 5 km s$^{-1}$ for model 2, respectively. This is somewhat lower than the maximum rotation velocities found in the simulation data (see bottom row of Fig.~\ref{fig:simulation}), which is expected due to the averaging process along the line of site.

The selected transitions shown in Figs.~\ref{fig:intensity} and~\ref{fig:intensity2} feature the brightest emission and strongest contrast between the disc and the ambient medium. For most of the remaining transitions (see Table~\ref{tab:lines}), an elongated disc structure is clearly recognisable, although for some of the lines the emission/contrast from the discs is much weaker. We emphasise that only for the transition H$^{13}$CO$^+$ $J$ = 1--0 the disc in the low-mass model 2 is not detectable.

\subsubsection{Impact of different transition energy levels}
\label{sec:J}

\begin{figure}
\includegraphics[width=\linewidth]{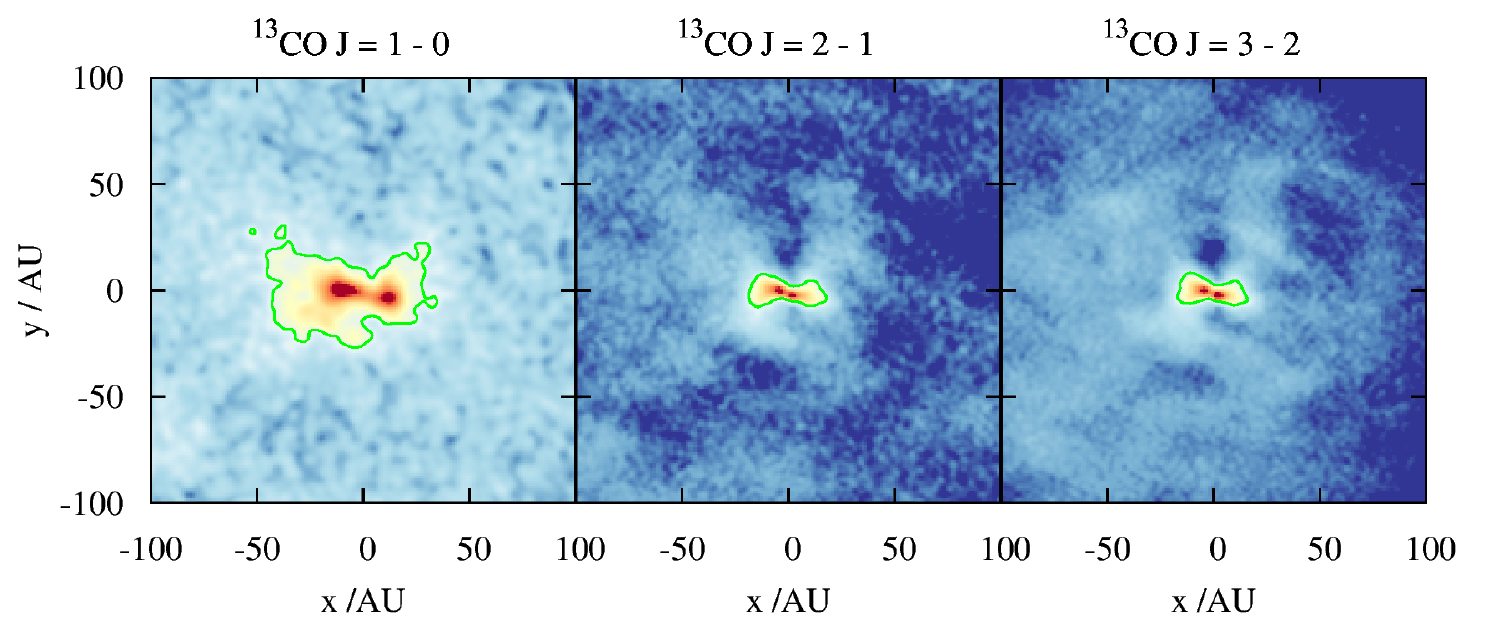}
\caption{Integrated intensity of $^{13}$CO for the transitions with $J$ = 1--0, 2--1, and 3--2 (from left to right). Increasing $J$ results in a more compact emission map as well as a higher peak brightness. The green line shows the contour where the intensity drops below 30 percent of the peak intensity.}
\label{fig:trans}
\end{figure}

So far, the observations probe only the inner parts of the discs. For this reason we investigate how the observed emission changes when considering transitions with different energy levels of a given molecule. We do this exemplarily for the $^{13}$CO lines in model 1 in Fig.~\ref{fig:trans}, where we show the integrated intensity for the transitions with $J$ = 1--0, 2--1, and 3--2.

Some characteristic results can be inferred from Fig.~\ref{fig:trans}. First, the emission tends to become more compact as $J$ increases. This is indicated by the green line showing the contour at which the intensity drops below 30 percent of the peak intensity: For $J$ = 1--0 the emission extends over radii of about 50 AU, which is closer to but still somewhat smaller than the radii of $\sim$ 70 AU over which the rotation is Keplerian (see bottom row of Fig.~\ref{fig:simulation}). For $J$ = 3--2, however, only the innermost disc region is probed.  This is due to the fact that higher $J$-transitions in general tend to probe warmer gas. Since the simulations show an increasing gas temperature towards the centre of the disc, higher $J$-transitions are thus probing smaller and more deeply embedded parts of the disc. Furthermore, it can be seen that the maximum emission increases with increasing $J$, as well as the contrast between the emission coming from the disc and the ambient medium. We emphasise that these results also hold for model 2 as well other considered species.

\subsubsection{Anisotropic accretion}
\label{sec:extend}

As discussed in \citet{Seifried15}, accretion onto the protostellar discs in the Class 0 stage seems to appear in a highly anisotropic manner on scales of a few 100 AU to 1000 AU. On larger scales (0.1 -- 1 pc) similar anisotropic accretion flows were observed in the case of massive protostars \citep{Beltran14,Liu15}. It is therefore interesting to see whether ALMA is able to detect signs of anisotropic accretion even on scales of 100 -- 1000 AU reported in \citet{Seifried15}.

\begin{figure}
 \includegraphics[width=\linewidth]{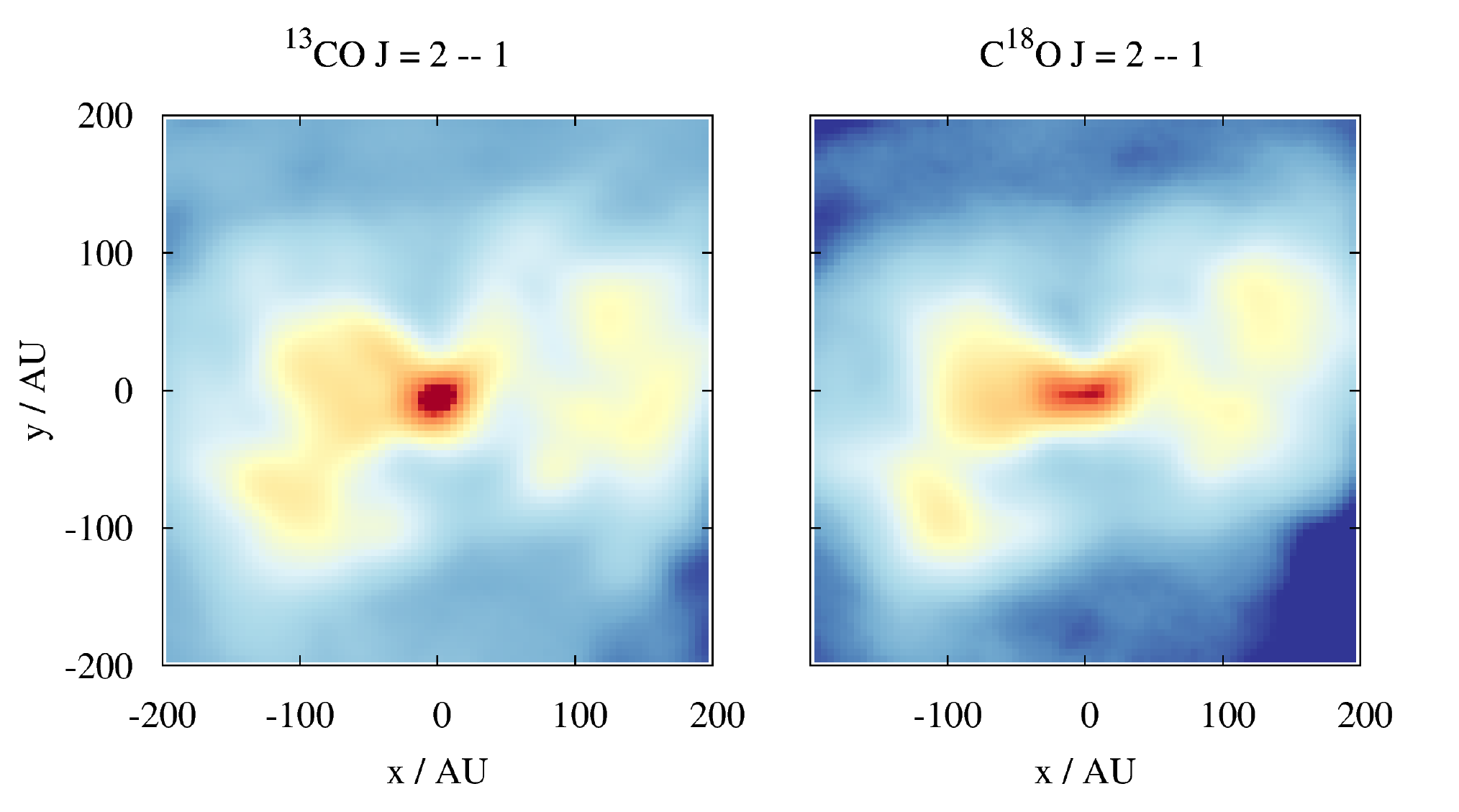}
 \caption{Intensities of $^{13}$CO $J$ = 2--1 (left) and C$^{18}$O $J$ = 2--1 (right) integrated over channels between --15 to --5 km s$^{-1}$ and 5 to 15 km s$^{-1}$ using a spatial resolution of 0.2'' (see text). The emission is more extended and shows some asymmetry.}
\label{fig:intensity_wide}
\end{figure}

As noted before, the disc emission (Figs.~\ref{fig:intensity} and~\ref{fig:intensity2}) is dominated by the velocity channels close to the systemic velocity\footnote{We note that the extended emission around the systemic velocity is filtered out by the interferometric observations. This, however, does not affect the detection of the disk which emits at larger velocities with respect to the systemic velocity (see Section~\ref{sec:pvdiag}). Moreover, in cases where the $^{13}$CO emission is abundant and optically thick, the opacity effects essentially dominate the emission at the systemic velocities still allowing for the detection of the disk at large velocities.}. Therefore, by visually inspecting the different channel maps of model 1, we can identify the channels which contain a more extended emission. In Fig.~\ref{fig:intensity_wide} we show the edge-on emission maps for $^{13}$CO $J$ = 2--1 and C$^{18}$O $J$ = 2--1 of model 1, which are integrated  from -- 15 to -- 5 km s$^{-1}$ and 5 to 15 km s$^{-1}$ and exclude the channels within $\pm$ 5 km s$^{-1}$ around the systemic velocity (bottom row). We emphasise that here we took the original high-resolution ALMA raw data and cleaned it under the assumption of a beam size of 0.2'' (note that at 150 pc a resolution of 0.2'' corresponds to a physical scale of 30 pc). Though this subsequent reduction of resolution somewhat reduces the overall intensity, we thus avoid to filter out the larger-scale emission -- an effect intrinsic to interferometric observations. When comparing with Fig.~\ref{fig:intensity}, one finds  that the emission is more diffuse and extends to significantly larger distances than the emission obtained by integrating over all channels. Moreover, the emission is markedly asymmetric with some prominent emission at the lower left of the disc. This asymmetric emission corresponds nicely to the results shown in Fig. 1 of \citep{Seifried15}, where we showed that the formation of stars is strongly affected by turbulence over a wide range of physical scales. It also shows that ALMA should be able to detect the continuation of the anisotropic accretion flow detected on larger scales \citep{Beltran14,Liu15}. A more detailed analysis of the observable signatures of this anisotropic accretion mode will be postponed to a subsequent paper.
For model 2 (not considered here) the emission remains rather smooth up to distances of 1000 AU and we do not detect any signs of anisotropic accretion.

\subsection{PV diagram and protostellar masses}
\label{sec:pvdiag}

Next, we investigate the dynamics of the discs by means of position-velocity (PV) diagrams. This allows us to test whether ALMA is able to trace Keplerian rotation signatures in protostellar discs around Class 0 objects. Furthermore, -- in case this is \mbox{possible --} we can verify how reliable protostellar masses are, which are obtained from fitting Keplerian profiles to the PV diagrams.

In order to do so, we construct the PV diagrams along the horizontal direction through the centre of the disc, taking into account pixels in a slit with a (vertical) height of about 6 times the beam size, which corresponds to a physical height of about 18 AU. In Fig.~\ref{fig:PV_diag} we show the PV diagram for the $^{13}$CO $J$ = 2--1 line for both models. In both cases the PV diagram reveals a clear Keplerian rotation profile with a good signal-to-noise ratio, which shows that at a source distance of 150 pc with the chosen observational parameters Keplerian rotation should be detectable with ALMA in the Class 0 stage. Interestingly, the profile extends up to radii of 70 -- 100 AU, thus in good agreement with the simulation data (see bottom row Fig.~\ref{fig:simulation}) and more extended than the intensities shown in Figs.~\ref{fig:intensity} and~\ref{fig:intensity2}.

\subsubsection{Fitting the ``upper'' edge}
\label{sec:upper}

The PV diagrams can be used to estimate the mass of the central protostar. In the following we explain our approach to obtain the protostellar mass by fitting a Keplerian profile to the PV diagrams:
\begin{itemize}
 \item First, we determine the noise level in the outer parts of the PV diagram where no emission is expected.
 \item In the next step, for each $x$-position, we determine the maximum rotation velocity $v_\rmn{rot,max}$. For this we only consider the upper left and the lower right quadrant of the PV diagram, i.e. those two quadrants in which most of the emission comes from. 
 \item Upper left quadrant:
 \begin{enumerate}
  \item For any given radius $x$ we start at the pixel with the maximum velocity of +60 km/s and determine the emission strength at this pixel.
  \item If the emission is lower than a certain threshold -- here 5 times the noise level determined in the beginning -- we go to the next lower velocity pixel.
  \item We repeat step (i) and (ii) until we reach the \textit{first} pixel where the emission exceeds the threshold. The velocity in this pixel gives the maximum rotation velocity $v_\rmn{rot,max}$ at the chosen radius $x$.
  \item We repeat this steps (i) -- (iii) for each radius and can thus identify the ``upper'' edge  of the PV diagram. 
  \end{enumerate}
 \item For the lower right quadrant the entire procedure (i) -- (iv) is identical except that for each radius we start at the channel with the minimum velocity, i.e. at --60 km/s and increase the velocity step by step until the intensity exceeds the threshold.
 \item We emphasise that in case for a given radius $x$ we do not find a pixel with emission above the threshold in the upper left or lower right quadrant we omit this data point.
\end{itemize}

\begin{figure}
\includegraphics[width=\linewidth]{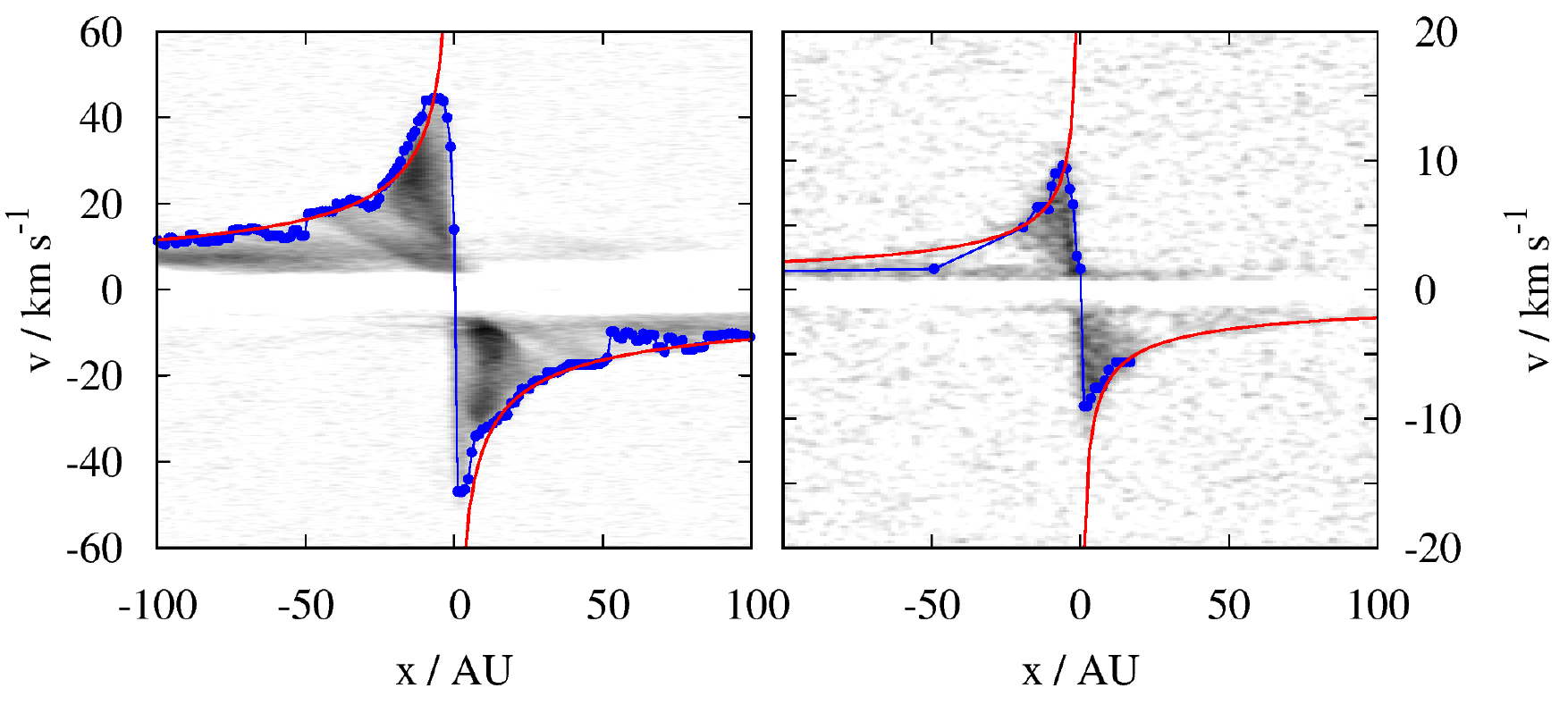}
\caption{PV diagram of model 1 (left) and 2 (right) for the $^{13}$CO $J$ = 2--1 line emission. The blue curve represent the maximum rotation velocity identified by our automatised approach (see text). The red curve represents the best fit of a Keplerian rotation profile to the blue curve.}
\label{fig:PV_diag}
\end{figure}

With this procedure we obtain $v_\rmn{rot,max}$ as a function of $x$. In Figure~\ref{fig:PV_diag}, we show $v_\rmn{rot,max}$ for the $^{13}$CO $J$ = 2--1 transition (blue curve). For model 1 the approach works fine up to large radii (left panel) whereas for model 2, in the outer parts, no maximum rotation velocity is identified. For both models, in the innermost part, $v_\rmn{rot,max}$ starts to decrease with decreasing radius, thus not representing a Keplerian profile any more, which is why we exclude the inner $\pm$ 5 AU from the fit.

Next, we fit a Keplerian profile $v(x) = \sqrt{G M / x}$ to the $v_\rmn{rot,max}$-curve obtained, which provides us the protostellar mass $M$. We show the fit (red line) to the curve of $v_\rmn{rot,max}$ in Figure~\ref{fig:PV_diag}. For model 1 we obtain a central mass of 15.0 $\pm$ 0.4 M$_{\sun}$, for model 2 a mass of 0.53 $\pm$ 0.03 M$_{\sun}$. This is in excellent agreement with the total mass of 15.3 M$_{\sun}$ of all protostars\footnote{Note that the protostar, which is somewhat off from the disc centre (see bottom left panel of Fig.~\ref{fig:simulation}), has a mass of less than 1 M$_{\sun}$, which is why to first order we here can assume that the entire protostellar mass is concentrated at the disc centre.} in the disc of model 1, as well as the protostellar mass of 0.62 M$_{\sun}$ in model 2.

We apply this routine to all models in which the disc is observed edge-on (see Section~\ref{sec:inclination} for inclination effects). The masses and standard deviation obtained by the fit for the different models and line transitions are listed in Table~\ref{tab:fits}.
\begin{table*}
   \caption{Overview of the protostellar masses including the standard deviation of the edge-on observed discs in the models 1 and 2 obtained for different lines by fitting a Keplerian profiles to the rotation velocity profiles obtained from the synthetic observations. Numbers in bracket denote values where the fit is rather poor.}
 \label{tab:fits}
  \begin{tabular}{cccccc}
  \hline
  & \multicolumn{5}{c}{masses for model 1 in M$_{\sun}$ (actual mass: 15.3 M$_{\sun}$)} \\
  \hline
  transition & $^{13}$CO 	& C$^{18}$O 		& HCO$^+$ 		& H$^{13}$CO$^+$ 	& N$_2$H$^+$ \\
  \hline
  $J$ = 1--0 & 13.0 $\pm$ 0.3	& 11.1 $\pm$ 0.5	& 14.8 $\pm$ 0.3	& No fit 		& 11.1 $\pm$ 0.5 \\
  $J$ = 2--1 & 15.0 $\pm$ 0.4	& 15.0 $\pm$ 0.2	& --- & ---& ---\\
  $J$ = 3--2 & 14.8 $\pm$ 0.4	& 14.9 $\pm$ 0.3	& 15.5 $\pm$ 0.3	& 13.1 $\pm$ 0.5	& 14.4 $\pm$ 0.4 \\
  $J$ = 4--3 & ---		& ---			& 15.4 $\pm$ 0.3	& 11.8 $\pm$ 0.6	& 13.5 $\pm$ 0.5 \\
  \hline
  \hline
   & \multicolumn{5}{c}{masses for model 2 in M$_{\sun}$ (actual mass: 0.62 M$_{\sun}$)} \\
  \hline
  $J$ = 1--0 & 0.48 $\pm$ 0.03	& 0.48 $\pm$ 0.02	& No fit		& No fit		& (0.43 $\pm$ 0.05) \\
  $J$ = 2--1 & 0.53 $\pm$ 0.03	& 0.53 $\pm$ 0.03	& --- & ---& --- \\
  $J$ = 3--2 & 0.47 $\pm$ 0.06	& 0.48 $\pm$ 0.05	& 0.47 $\pm$ 0.06	& 0.52 $\pm$ 0.02	& 0.49 $\pm$ 0.06 \\
  $J$ = 4--3 & ---		& ---			& 0.40 $\pm$ 0.06	& 0.55 $\pm$ 0.02	& 0.38 $\pm$ 0.05\\
 \hline
  \end{tabular}
\hspace{-0.3cm}
 \begin{tabular}{ccc}
  \hline
  \multicolumn{3}{c}{model 1} \\
  \hline
  frequency & o-H$_2$CO & p-H$_2$CO \\
  \hline
  226 GHz & 15.5 $\pm$ 0.3 & --- \\
  146 GHz & --- & 15.0 $\pm$ 0.3\\
  218 GHz & --- &  15.6 $\pm$ 0.3 \\
  363 GHz & --- &  15.7 $\pm$ 0.3\\
  \hline
  \hline
  \multicolumn{3}{c}{model 2} \\
  \hline
  226 GHz & 0.53 $\pm$ 0.05 & --- \\
  146 GHz & --- & (0.38 $\pm$ 0.04) \\
  218 GHz & --- & 0.52 $\pm$ 0.40 \\
  363 GHz & --- & 0.59 $\pm$ 0.03 \\
  \hline
 \end{tabular}
\end{table*}
For model 1, for all but one line, the derived protostellar masses agree within \mbox{$\sim$ 25 percent} of the actual protostellar mass, for some even within \mbox{10 percent}. For H$^{13}$CO$^+$ $J$ = 1--0 no mass could be determined from the PV diagram, which is due to the fact that H$^{13}$CO$^+$ has a low abundance, which in turn results in a relatively noisy PV diagram.

For the low-mass model 2, in particular the determination of the protostellar mass from the $J$ = 1--0 transitions is problematic. This difficulty arises from the weak emission and the corresponding low signal-to-noise ratio in the emission maps as already shown in Fig.~\ref{fig:intensity2}. By inspecting the fit results by eye we find that in particular for N$_2$H$^+$ $J$ = 1--0 and H$_2$CO (146 GHz) the fits are rather poor, which is why we put the values in brackets. For HCO$^+$ $J$ = 1--0 and H$^{13}$CO$^+$ no protostellar masses could be determined although hints of a Keplerian rotation profile are recognisable in the PV diagrams. For the remaining lines, however, the fits give reasonable results with protostellar masses off by at most 35\% from the actual mass of 0.62 M$_{\sun}$.

However, in both models in most cases the actual masses are \textit{outside} the error limits of the observed masses -- in general somewhat too low-- although the differences are not significant. This points to a systematic uncertainty in the determination of the protostellar mass generally resulting in somewhat too low masses.

To summarize, we find that for the chosen observational parameters of 5 h observing time, a resolution of 0.02'' and a pwv of 0.5 mm, ALMA should be able to detect Keplerian rotation profiles around Class 0 protostellar objects up to radii of 100 AU. Furthermore, predicted protostellar masses show a reasonable accuracy in the case the discs are seen edge-on.

\subsection{Influence of the viewing angle}
\label{sec:inclination}

So far, we have only considered discs seen edge-on, which is the ideal case to study their dynamics. However, since such a configuration is rather unlikely, we investigate how an inclination with respect to the line-of-sight affects the identification of Keplerian profiles as well as the determination of protostellar masses.

\begin{figure*}
 \includegraphics[width=\linewidth]{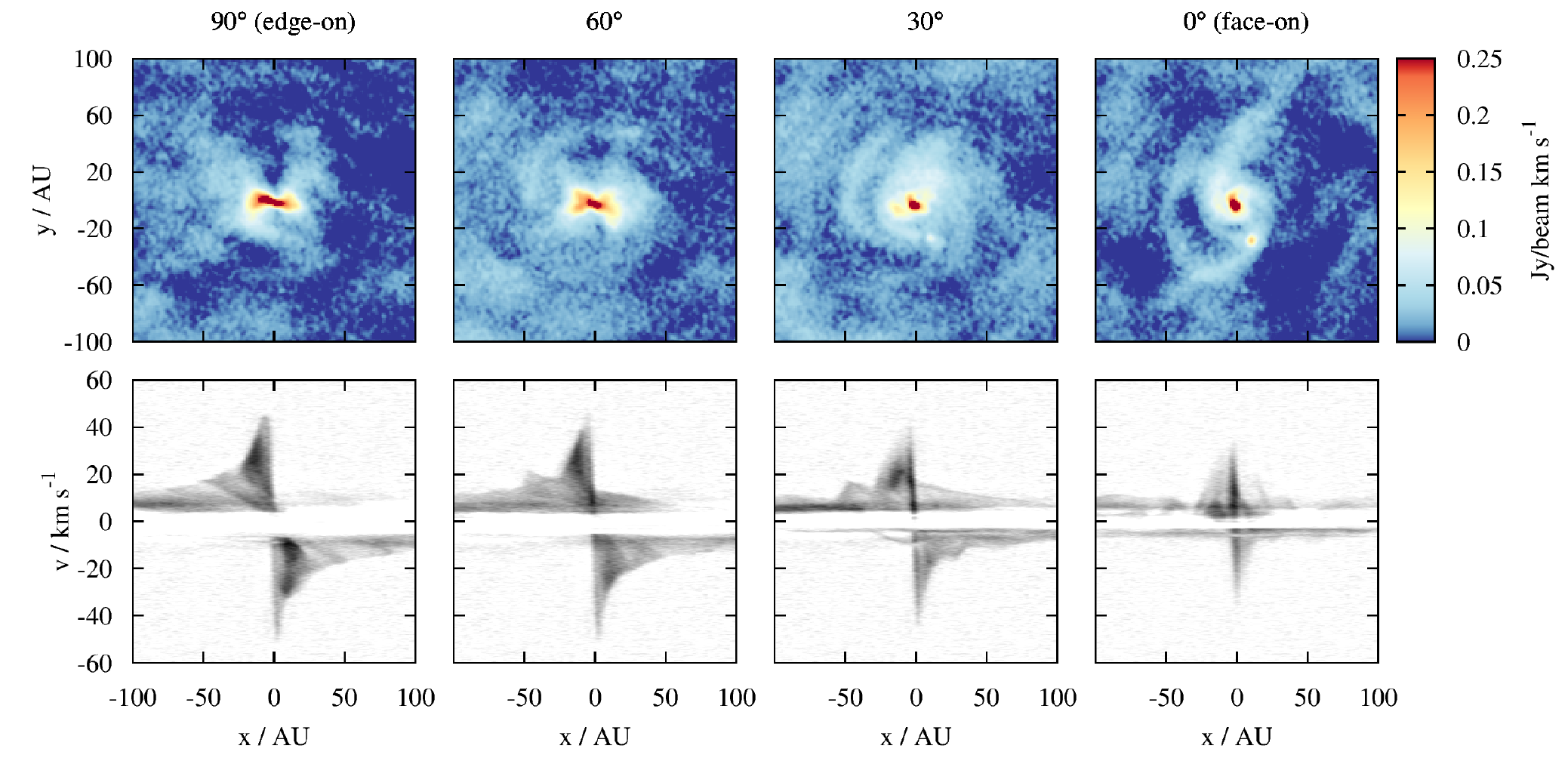}
 \caption{Integrated intensity (top row), and PV diagram (bottom row) for the $^{13}$CO $J$ = 2--1 line emission of model 1 for inclination angles of 90$^\circ$, 60$^\circ$, 30$^\circ$ and 0$^\circ$ (left to right) under optimal observing conditions (see Table~\ref{tab:casa}) at a distance of 150 pc.}
 \label{fig:incl} 
\end{figure*}

In Figure~\ref{fig:incl} we plot the 0th-order moment maps as well as the PV diagrams for model 1 for an inclination angle of 90$^\circ$, 60$^\circ$, 30$^\circ$ and 0$^\circ$, thus from edge-on to face-on, for our fiducial case of the $^{13}$CO $J$ = 2--1 line.
Up to an inclination angle of 30$^{\circ}$, a Keplerian velocity profile is recognisable in the PV diagram. With the method described in the previous section, we obtain protostellar masses of 15.0, 12.8, and 7.9 $M_{\sun}$ for inclination angles of 90$^{\circ}$, 60$^{\circ}$, and 30$^{\circ}$, respectively. For the face-on disc we cannot not apply any fit.

Since the mass obtained by the fit should scale with $v_\rmn{rot,max}^2$ and since $v_\rmn{rot,max}$ decreases with the inclination angle $\theta$ as cos(90$^\circ$ - $\theta$), we would expect the fitted masses to scale as cos$^2$(90$^\circ$ - $\theta)$. Hence, the obtained masses for $\theta$ =  60$^{\circ}$ and 30$^{\circ}$ should be smaller than the mass of the edge-on case by a factor of 0.75 and 0.25, respectively. However, the observed masses are clearly larger than the theoretically expected values (see Table~\ref{tab:fitsincl}) and do not agree within their error limits. For $\theta = 30^{\circ}$, the obtained masses exceed the theoretical value by a factor of more than 2. Considering the bottom panels of Fig.~\ref{fig:simulation} shows that the gas in the discs have considerable radial motions (despite being smaller than the rotational motions). When increasing the inclination, these radial motions progressively contribute to the line-of-sight velocity, hence also increasing the apparent rotation velocity. This in turn results in systematically higher masses, in particular for large inclinations. Furthermore, the PV-diagrams for $\theta$ =  60$^{\circ}$ and 30$^{\circ}$ reveal some asymmetry between the red- and blue-shifted part. This could be due to the highly anisotropic accretion onto the discs discussed in Section~\ref{sec:extend} \citep[see also][]{Seifried15}: Infalling gas close to the disc, which is located in the fore- or background, might contribute to the emission at high velocity channels. This gas further complicates the determination of protostellar masses for large inclinations.

\begin{figure*}
 \includegraphics[width=0.9\linewidth]{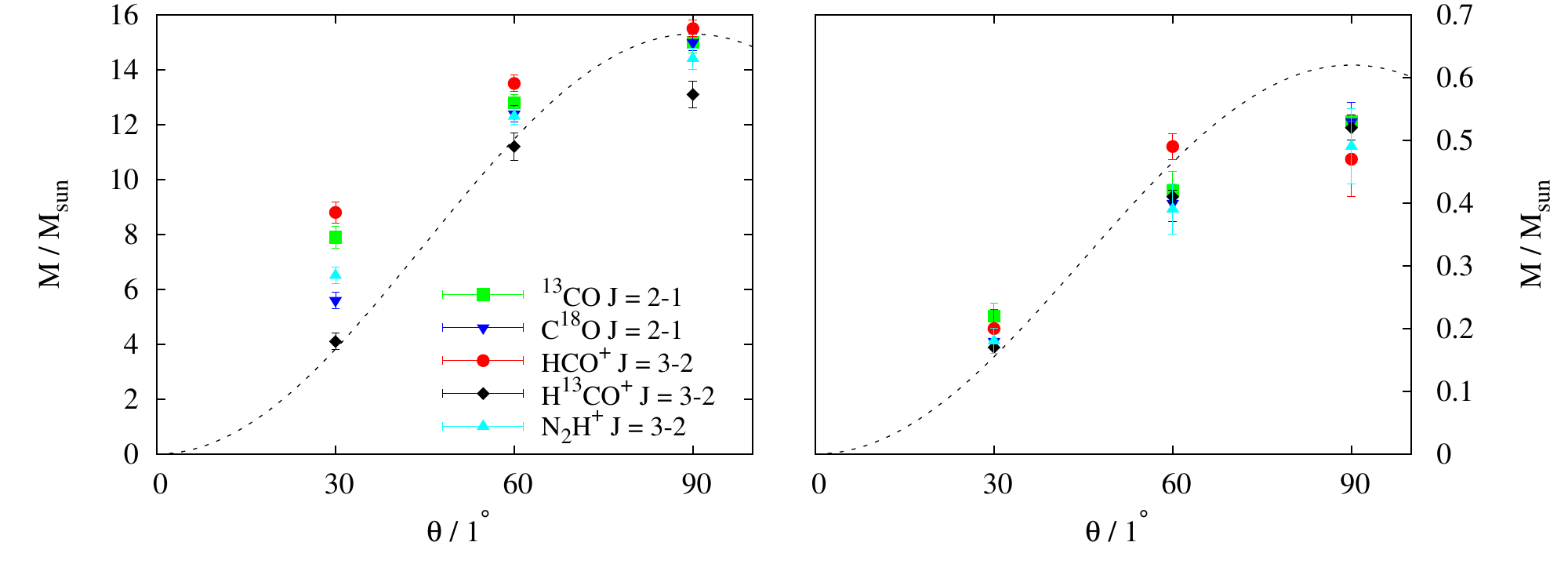}
 \caption{Protostellar masses obtained from the PV diagrams including the standard deviation of model 1 (left) and 2 (right) for different lines and inclination angles $\theta$. The black dashed line shows the theoretical expectation (see text).}
 \label{fig:masses} 
\end{figure*}

We repeat the above analysis for the lines C$^{18}$O $J$ = 2--1, HCO$^+$ $J$ = 3--2, H$^{13}$CO$^+$ $J$ = 3--2, and N$_2$H$^+$ $J$ = 3--2, for both model 1 and 2 (for model 2 $^{13}$CO $J$ = 2--1 is considered as well) and list the results in Table~\ref{tab:fitsincl}. In addition, in Fig.~\ref{fig:masses} we show the obtained protostellar masses for the different lines and inclination angles and compare them to the theoretical expectation (black dashed line).
\begin{table*}
   \caption{Overview of the protostellar masses of model 1 and 2 obtained by fitting Keplerian profiles to the rotation velocity profiles obtained from the synthetic observations for different lines and inclination angles including the standard deviation.}
 \label{tab:fitsincl}
 \begin{tabular}{cccccc}
  \hline
  & \multicolumn{5}{c}{masses for model 1 in M$_{\sun}$ (expected mass: 90$^{\circ}$: 15.3 M$_{\sun}$, 60$^{\circ}$: 11.4 M$_{\sun}$, 30$^{\circ}$: 3.8 M$_{\sun}$)} \\
  \hline
  inclination, line & $^{13}$CO $J$ = 2--1 	& C$^{18}$O $J$ = 2--1 	& HCO$^+$ $J$ = 3--2 	& H$^{13}$CO$^+$ $J$ = 3--2 & N$_2$H$^+$ $J$ = 3--2 \\
  \hline
  30$^{\circ}$ & 7.9 $\pm$ 0.4		& 5.6 $\pm$ 0.3		& 8.8 $\pm$ 0.4		& 4.1 $\pm$ 0.3		& 6.5 $\pm$ 0.3 \\
  60$^{\circ}$ & 12.8 $\pm$ 0.3		& 12.4 $\pm$ 0.3	& 13.5 $\pm$ 0.3	& 11.2 $\pm$ 0.5	& 12.3 $\pm$ 0.3 \\
  90$^{\circ}$ & 15.0 $\pm$ 0.4		& 15.0 $\pm$ 0.2	& 15.5 $\pm$ 0.3	& 13.1 $\pm$ 0.5	& 14.4 $\pm$ 0.4 \\
  \hline
  \hline
   & \multicolumn{5}{c}{masses for model 2 in M$_{\sun}$ (expected mass: 90$^{\circ}$: 0.62 M$_{\sun}$, 60$^{\circ}$: 0.47 M$_{\sun}$, 30$^{\circ}$: 0.16 M$_{\sun}$)} \\
  \hline
  30$^{\circ}$ & 0.22 $\pm$ 0.02	& 0.18 $\pm$ 0.02	& 0.20 $\pm$ 0.03	& 0.17 $\pm$ 0.01	& 0.18 $\pm$ 0.02 \\
  60$^{\circ}$ & 0.42 $\pm$ 0.03	& 0.40 $\pm$ 0.03	& 0.49 $\pm$ 0.02	& 0.41 $\pm$ 0.01	& 0.39 $\pm$ 0.04 \\
  90$^{\circ}$ & 0.53 $\pm$ 0.03	& 0.53 $\pm$ 0.03	& 0.47 $\pm$ 0.06	& 0.52 $\pm$ 0.02	& 0.49 $\pm$ 0.06 \\
 \hline
  \end{tabular}
\end{table*}
As discussed in Section~\ref{sec:pvdiag}, the masses for edge-on discs are in general somewhat too low. For a viewing angle of 60$^\circ$ the values scatter around the theoretically expected value: the values of model 1 appear to be somewhat too high, whereas the values of model 2 are somewhat too low. At an inclination angle of 30$^{\circ}$, for both models, the protostellar masses inferred from the fitting routine are higher than the theoretically expected value and also outside the error limits of the fits. This again supports our assumption that infall motions in the discs and potentially also in the fore- or background are erroneously interpreted as rotation velocities.

In general, however, we find that even for a strongly inclined disc ($\sim$ 30$^{\circ}$) we obtain protostellar masses, which are within a range of a few 10\% around the actual mass. Finally, we note that similar results are also found for the remaining lines listed in Table~\ref{tab:lines} but not in Table~\ref{tab:fitsincl}.

\subsection{Disk fragmentation}
\label{sec:frag}

As shown in the left panel of Fig.~\ref{fig:simulation}, the disc in model 1 is fragmented after 35 kyr, while the disc in model 2 remains gravitationally stable. In the following, we study whether such fragmentation features can be probed with ALMA with the current observational parameters for face-on discs. Considering the top right panel of Fig.~\ref{fig:incl}, we can indeed identify such signatures of protostellar disc fragmentation, namely  two spiral arms extending in the vertical direction as well as two bright clumps with one being located in the lower spiral arm.

\begin{figure*}
 \includegraphics[width=\linewidth]{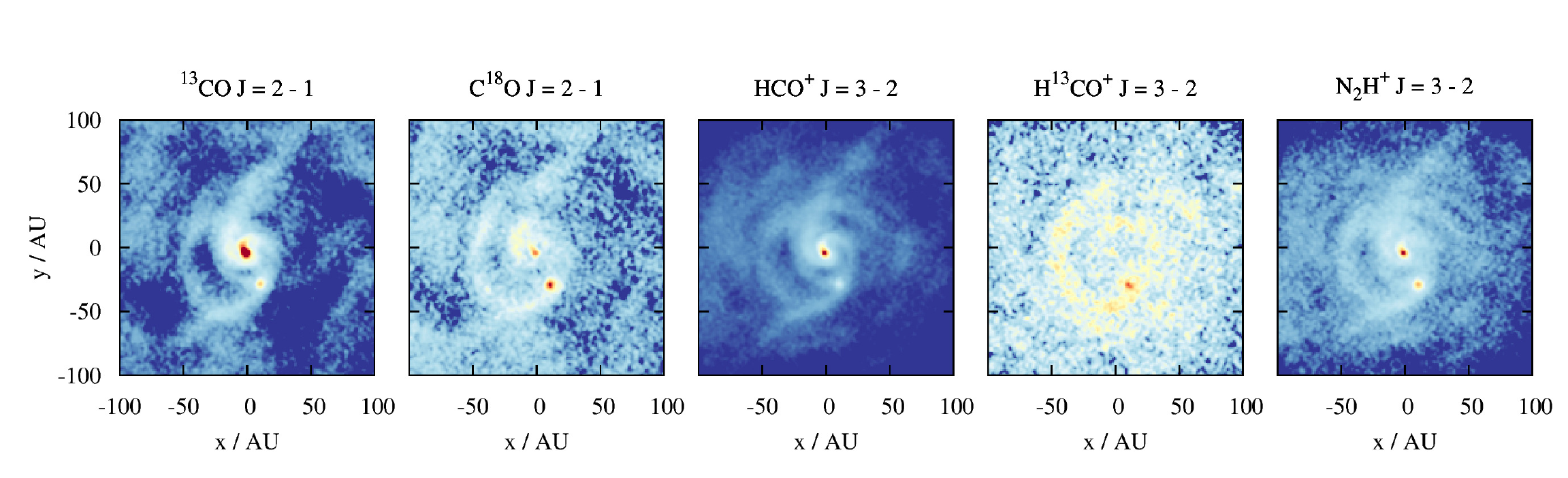}
 \caption{Face-on view of the disc of model 1 for the same lines as in Fig.~\ref{fig:intensity} (except for p-H$_2$CO (218 GHz)). Signs of fragmentation and spiral arms are visible for all lines considered.}
 \label{fig:frag}
\end{figure*}

In Fig.~\ref{fig:frag} we show the face-on view of the disc in model 1 for $^{13}$CO $J$ = 2--1, C$^{18}$O $J$ = 2--1, HCO$^+$ $J$ = 3--2, H$^{13}$CO$^+$ $J$ = 3--2, and N$_2$H$^+$ $J$ = 3--2 (the same line transitions as in Fig.~\ref{fig:intensity}). For all transitions the fragmentation and spiral structure is visible in the ALMA image. For H$^{13}$CO$^+$ the emission is relatively diffuse and the signatures of fragmentation are rather weak, which is most likely due to its low fractional abundance.

Since we did not include radiative feedback from the protostars the degree of fragmentation might somewhat be overestimated. Nevertheless, our results indicate that if discs fragment in the Class 0 stage this fragmentation should be observable with ALMA in various molecular lines for good observing conditions and face-on discs.

\section{Influence of observing conditions}
\label{sec:conditions}

\subsection{Observing time and weather conditions}

So far we considered optimal observing conditions, but these do not always prevail. Therefore we redo our analysis for both models in $^{13}$CO $J$ = 2--1 and HCO$^+$ $J$ = 3--2 and consider moderate observing conditions with an observing time of 2 h and a pwv of 1 mm; the resolution of 0.02'' remains.

\begin{figure*}
 \includegraphics[width=\linewidth]{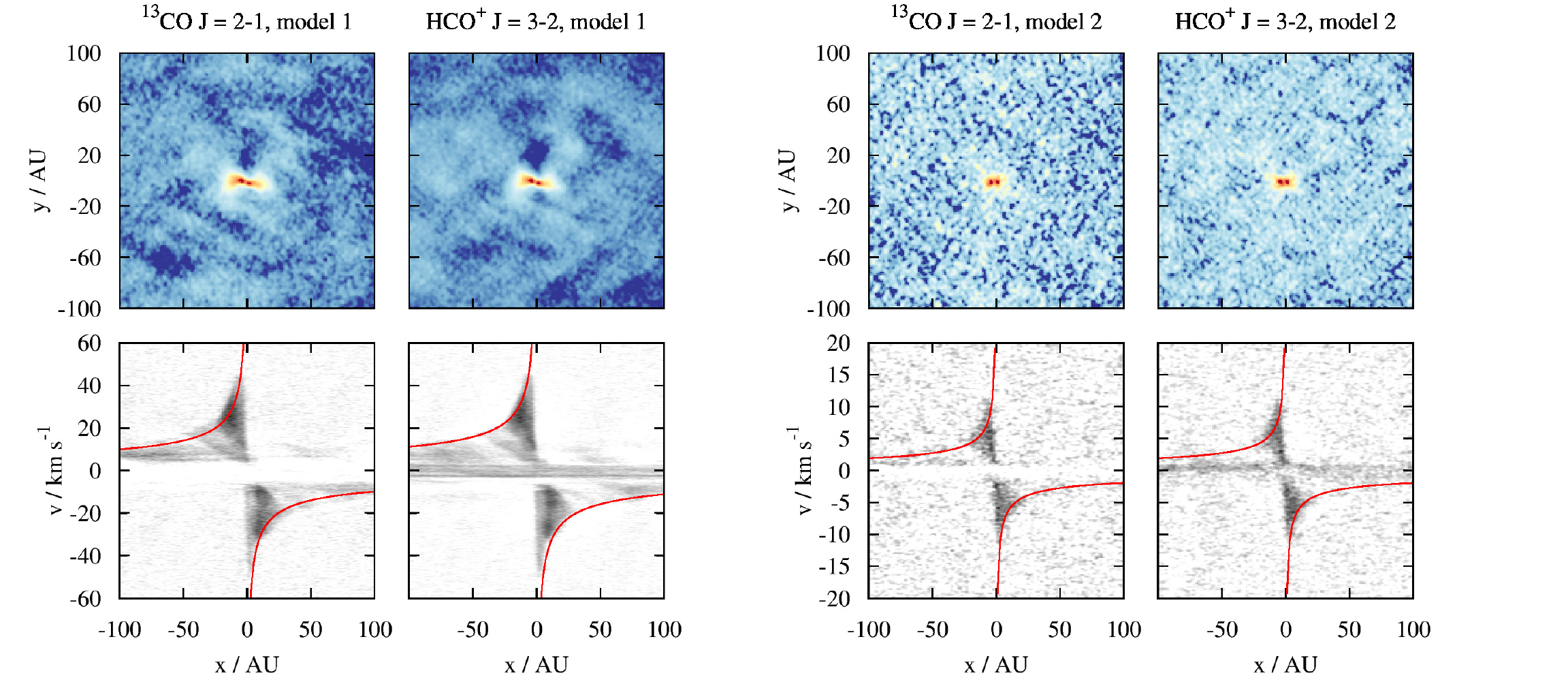}
 \caption{Integrated intensities (top row) and PV diagrams (bottom row) for the lines $^{13}$CO $J$ = 2--1 and HCO$^+$ $J$ = 3--2 and both models 1 (left) and 2 (right) adopting a pwv of 1 mm and an observing time of 2 h. The red lines show the Keplerian rotation curve for the protostellar masses obtained by the ``upper'' edge fits.}
 \label{fig:badcond}
\end{figure*}

In Fig.~\ref{fig:badcond} we show the integrated intensity as well as the PV diagrams of the aforementioned lines. As can be seen, for the intermediate-mass model 1 the disc (top row) and the Keplerian rotation profile (bottom row) are clearly recognisable for both transitions. Fitting a Keplerian velocity profile, we obtain masses which are in reasonable agreement with the actual protostellar mass (see left part of  Table~\ref{tab:fitsbadincl}). The Keplerian rotation curve for the obtained protostellar masses is also shown (red lines), and appears to be in good agreement with the underlying PV diagrams.
For model 2, however, the less optimal observational conditions are somewhat more problematic: a Keplerian velocity profile is recognisable, however, only extending to a radius of about 25 AU, which is why we restrict the fit to this radius. Nonetheless, the fit gives reasonable results comparable to the actual mass of 0.62 M$_{\sun}$ (left part of Table~\ref{tab:fitsbadincl}). Also the Keplerian velocity profile for the actual central mass of 0.62 M$_{\sun}$ shown in the right panel of Fig.~\ref{fig:badcond} fits the observed rotation profile quite well.

\begin{figure}
 \includegraphics[width=\linewidth]{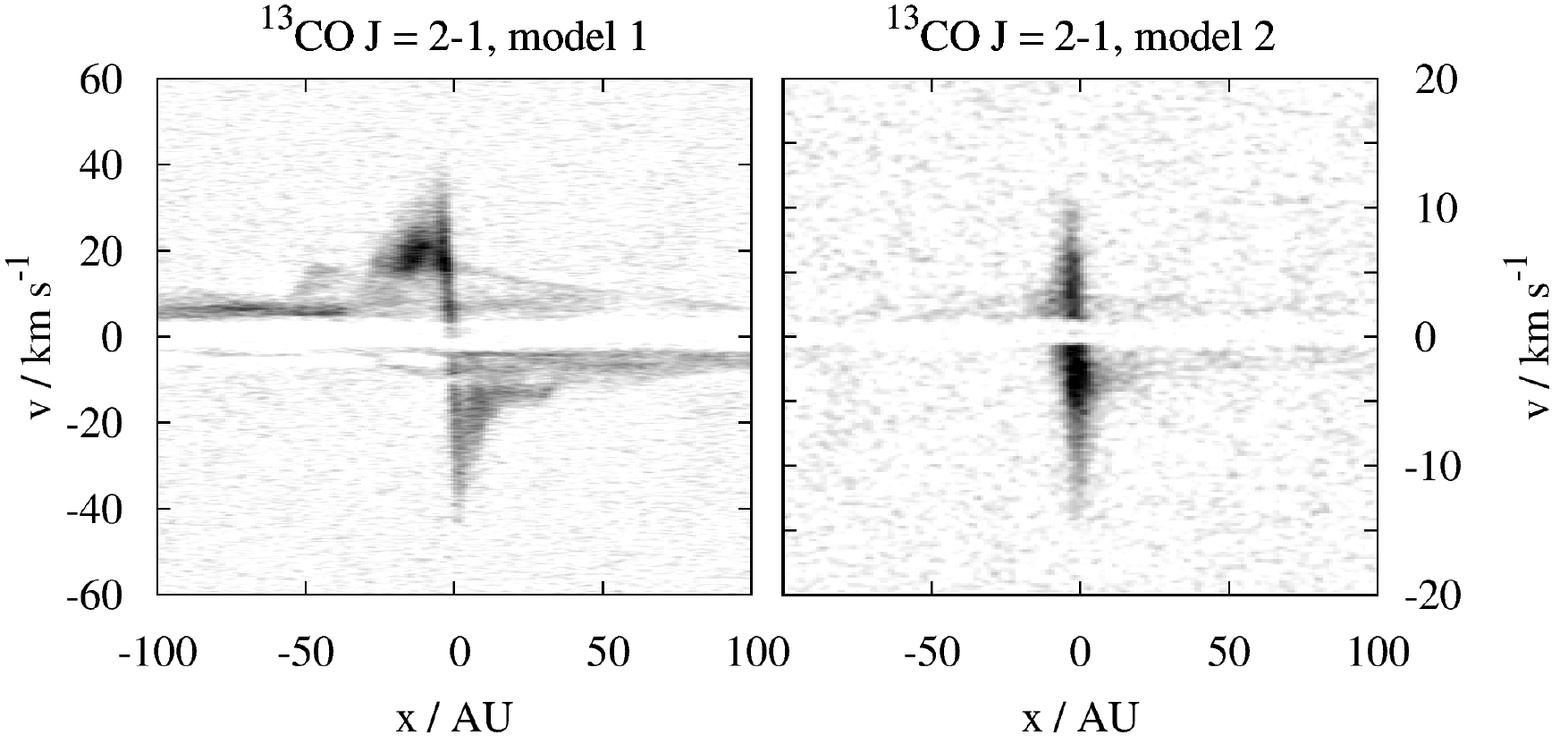}
 \caption{PV diagrams for the line $^{13}$CO $J$ = 2--1 of model 1 (left) and 2 (right) adopting a pwv of 1 mm, an observing time of 2 h, and an disc inclination of 30$^{\circ}$. No Keplerian profile is recognisable for model 2.}
 \label{fig:badcondincl}
\end{figure}

We extend the analysis with the aforementioned conditions for both lines and models assuming a disc inclination of 60$^{\circ}$ and 30$^{\circ}$ (see left part of Table~\ref{tab:fitsbadincl}). 
In model 1, for both models and lines considered, we are able to derive protostellar masses in rough agreement with the theoretically expected values of 11.4 and 0.47 M$_{\sun}$ for model 1 and 2, respectively. In general, they appear to be somewhat below the corresponding masses obtained for the synthetic observation under optimal conditions (see Table~\ref{tab:fitsincl}). For an inclination of 30$^{\circ}$, however, the situation is somewhat different. As can be seen in Fig.~\ref{fig:badcondincl}, where we show the PV diagrams of $^{13}$CO $J$ = 2--1 for both models, only for model 1 a Keplerian profile can be identified. This conclusion also holds for the PV diagrams constructed from the HCO$^+$ $J$ = 3--2 emission (not shown here). In general, however, for both the low- and intermediate-mass case, it should be possible with ALMA to determine the dynamics of Class 0 protostellar discs and the protostellar masses even under less optimal conditions.
\begin{table*}
   \caption{Overview of the protostellar masses of model 1 and 2 obtained by fitting Keplerian profiles to the rotation velocity profiles for less optimal conditions. In both cases in model 2 for an inclination of 30$^{\circ}$ no protostellar mass can be determined. Left: resolution of 0.02'' -- 0.05'', 2 h observing time, and a pwv of 1 mm. Right: resolution of 0.1'', 2 h observing time, and a pwv of 1 mm}
 \label{tab:fitsbadincl}
 \centering
 \begin{tabular}{ccc}
  \hline
  & \multicolumn{2}{c}{masses for model 1 in M$_{\sun}$} \\
  \hline
  inclination, line & $^{13}$CO $J$ = 2--1 	& HCO$^+$ $J$ = 3--2 	 \\
  \hline
  30$^{\circ}$ & 5.1 $\pm$ 0.2		& 6.0 $\pm$ 0.8 \\
  60$^{\circ}$ & 9.3 $\pm$ 0.4		& 12.1 $\pm$ 0.6 \\
  90$^{\circ}$ & 11.1 $\pm$ 0.4		& 14.0 $\pm$ 0.6 \\
  \hline
  \hline
   & \multicolumn{2}{c}{masses for model 2 in M$_{\sun}$} \\
  \hline
  60$^{\circ}$ & 0.34 $\pm$ 0.02 & 0.33 $\pm$ 0.03 \\
  90$^{\circ}$ & 0.42 $\pm$ 0.05 & 0.42 $\pm$ 0.05 \\
 \hline
 \end{tabular}
  \hspace{1cm}
 \begin{tabular}{ccc}
  \hline
  & \multicolumn{2}{c}{masses for model 1 in M$_{\sun}$} \\
  \hline
  inclination, line & $^{13}$CO $J$ = 2--1 	& HCO$^+$ $J$ = 3--2 	 \\
  \hline
  30$^{\circ}$ & 8.0 $\pm$ 0.4		& 9.7 $\pm$ 0.4 \\
  60$^{\circ}$ & 12.8 $\pm$ 0.6		& 14.4 $\pm$ 0.6 \\
  90$^{\circ}$ & 14.6 $\pm$ 0.3		& 16.2 $\pm$ 0.3 \\
  \hline
  \hline
   & \multicolumn{2}{c}{masses for model 2 in M$_{\sun}$} \\
  \hline
  60$^{\circ}$ & 0.51 $\pm$ 0.02 & 0.50 $\pm$ 0.02 \\
  90$^{\circ}$ & 0.57 $\pm$ 0.01 & 0.51 $\pm$ 0.02 \\
 \hline
  \end{tabular}
\end{table*}

\subsection{Resolution dependence and source distance}
\label{sec:distance}

\begin{figure*}
 \includegraphics[width=\linewidth]{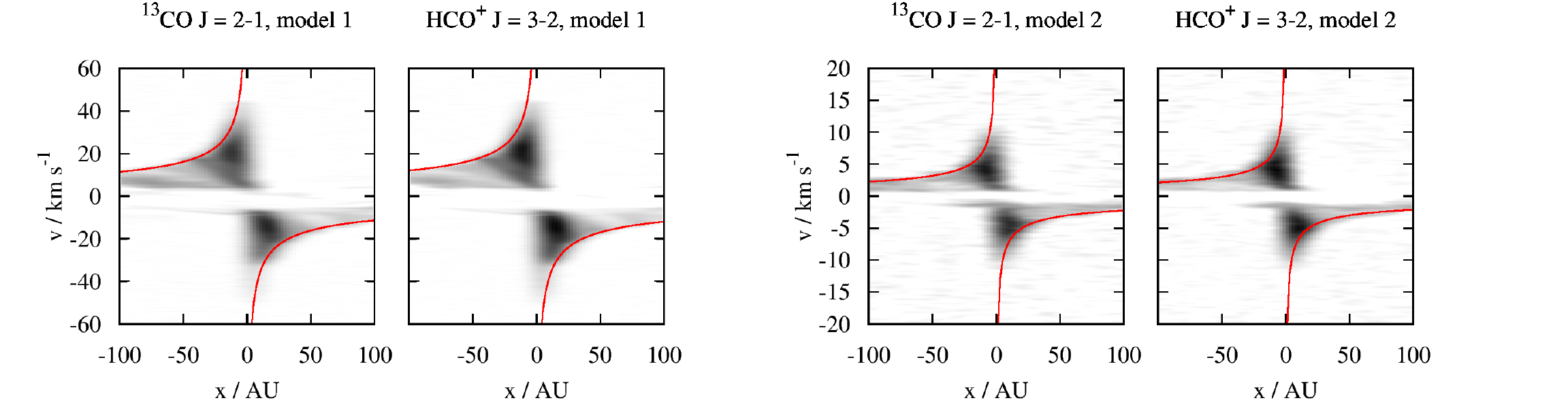}
 \caption{PV diagrams of the edge-on case for the lines $^{13}$CO $J$ = 2--1 and HCO$^+$ $J$ = 3--2 of model 1 (left) and 2 (right), for a resolution of 0.1'', a pwv of 1 mm, and an observing time of 2 h.}
 \label{fig:res}
\end{figure*}

In a second step, we investigate the influence of the resolution using the same lines as in the previous section. However, we now use a five times coarser resolution of 0.1'', and an observing time of 2 h and a pwv of 1 mm. With this resolution, which corresponds to 15 AU at an assumed distance of 150 pc, the integrated emission coming from the discs (not shown) is almost circular without any clear elongation. In Fig.~\ref{fig:res} we show the corresponding PV diagrams for both models in the edge-on case for $^{13}$CO $J$ = 2--1 and HCO$^+$ $J$ = 3--2. With the degraded resolution, Keplerian rotation profiles can still be seen for both models, although model 2 is more noisy.

We list the fitted protostellar masses in the right part of Table~\ref{tab:fitsbadincl}\footnote{Note that due to the low noise level, here we had to increase the threshold, for which we accept a pixel, to 20 times the noise level (see step (ii) in the approach described in Section~\ref{sec:upper}).} and plot the corresponding Keplerian rotation curves in Fig.~\ref{fig:res} . In general, we find the protostellar masses to be somewhat too high, although the difference is in general not more than $\sim$ 50\%.
We note that for an inclination of 30$^{\circ}$ only hints of a Keplerian profile for model 1 can be found whereas for model 2 no conclusions can be made about the dynamical state of the disc.

\begin{figure}
 \includegraphics[width=\linewidth]{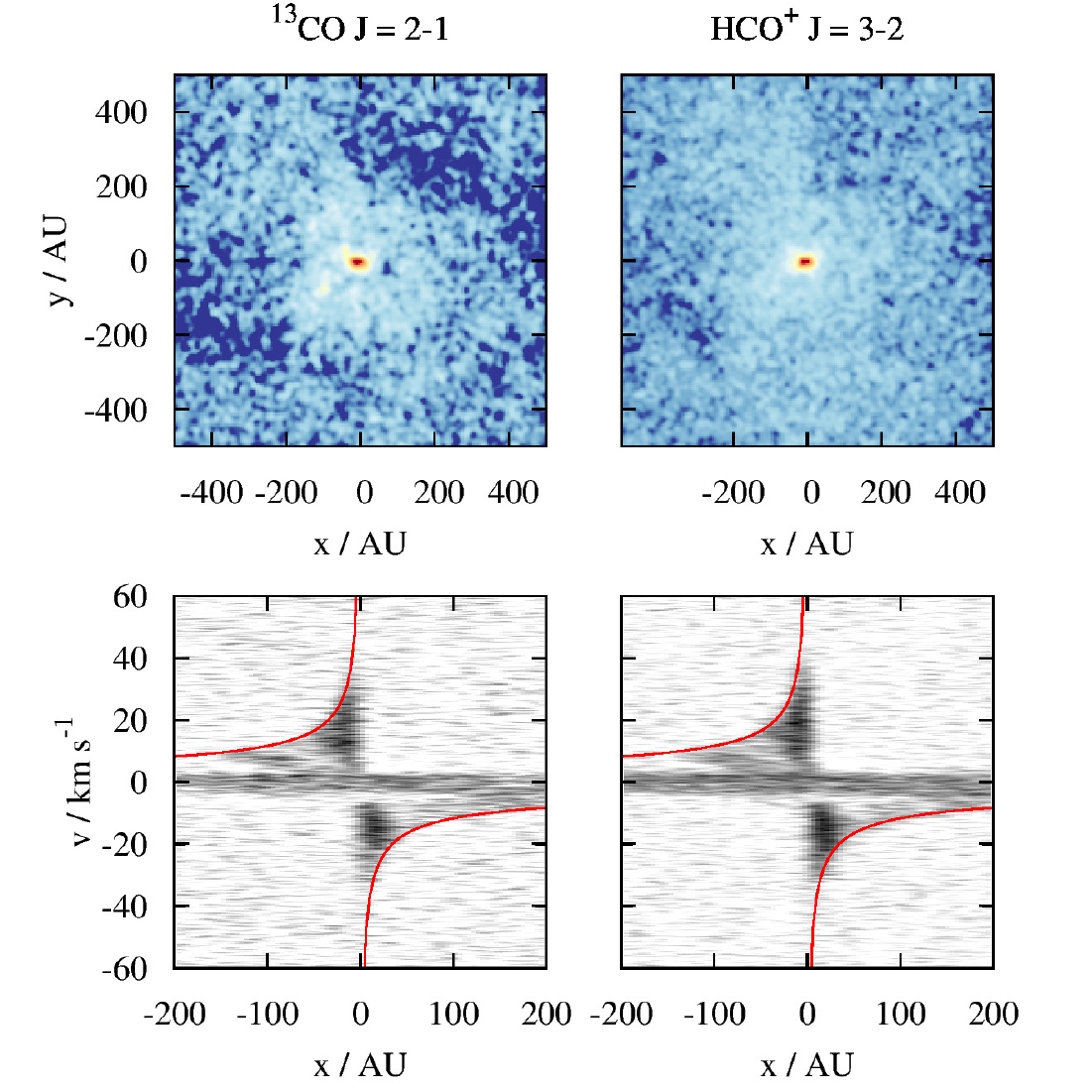}
 \caption{Integrated intensity (top) and PV diagram (bottom) for the transitions $^{13}$CO $J$ = 2--1 (left) and HCO$^{+}$ $J$ = 3--2 (right) in model 1 assuming a distance of 1 kpc. The red lines show the Keplerian rotation curve for a central mass of 15.3 M$_{\sun}$ inferred from the simulation.}
 \label{fig:distance}
\end{figure}

Finally, we investigate what happens if the disc of model 1 is located at a distance of 1 kpc, which corresponds to typical distances of intermediate- to high-mass star forming regions. In Fig.~\ref{fig:distance} we show the synthetic ALMA observations of $^{13}$CO $J$ = 2--1 and HCO$^{+}$ \mbox{$J$ = 3--2} assuming optimal observing conditions (a pwv of 0.5 mm and an observing time of 5 h).
The integrated emission (top row) does not reveal a clear elongated structure typical for discs. However, in the PV diagram (bottom row) a Keplerian rotation profile is clearly recognisable. Plotting the Keplerian rotation curve corresponding to a central mass of 15.3 M$_{\sun}$ inferred from the simulation (red lines), shows that it fits the observed PV diagram quite well. This indicates that with ALMA it should be possible to estimate protostellar masses even at a distance of 1 kpc. We note that fitting the ``upper'' edge results in protostellar masses around 7 M$_{\sun}$, which we attribute to the more noisy background in the PV diagrams.

Moreover, the results for inclination angles of 60$^{\circ}$ and 30$^{\circ}$ show that Keplerian profiles can be identified for both lines and angles considered, although for 30$^{\circ}$ the identification is somewhat more difficult. The results thus clearly demonstrate the ability of ALMA to probe the dynamics of even strongly inclined discs at source distance typical for intermediate- to high-mass star forming regions.

\subsection{Summary}

\begin{figure}
 \includegraphics[width=\linewidth]{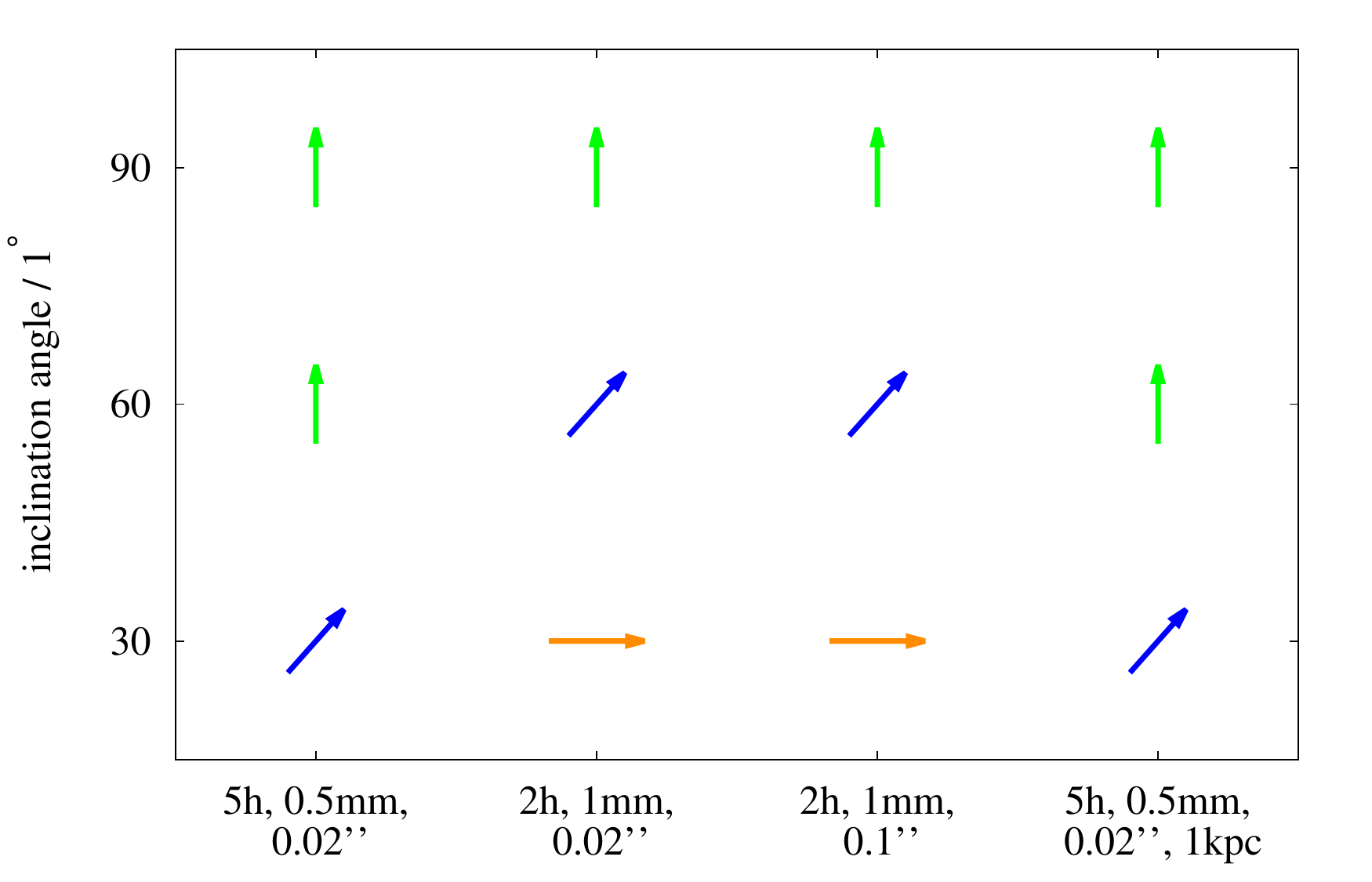}
 \caption{Impact of observing time, weather conditions, resolution, and source distance on the ability of ALMA to identify Keplerian motions for various disc inclination angles. Green up-vectors denote that Keplerian profiles can easily be identified, blue vectors to the top right indicate some minor restrictions. A horizontal orange vector indicates that only for model 1 (intermediate-mass protostar) Keplerian rotation can be identified, whereas for the low-mass model 2 it cannot be identified.}
 \label{fig:overview}
\end{figure}

In Fig.~\ref{fig:overview} we summarize the results of this section. We show observing conditions on the $x$-axis and the inclination angle on the $y$-axis. The arrows indicate whether discs in Keplerian rotation can be identified from the observations.
In general, ALMA should be able to detect discs and to identify Keplerian rotation for moderately inclined disc (90$^{\circ}$ and 60$^{\circ}$) at a source distance of 150 pc even in the case of moderate weather conditions (1 mm pwv), an observing time of 2 h, and a resolution of 0.1''. For strongly inclined discs (30$^{\circ}$) the identification of Keplerian rotation signatures is more problematic in particular for discs around low-mass protostars. Moreover, our results show that, when using the full capacity of ALMA, the dynamics of discs can be probed down to scales of $\sim$ 100 AU even at kpc distances.

\section{Discussion}
\label{sec:discussion}

\subsection{Sizes of Keplerian discs}

In Section~\ref{sec:results}, we show that the diameter of the disc in model 2 as extracted from the integrated emission maps (Fig.~\ref{fig:intensity2}) and the part of the PV diagram that is well fit by a Keplerian profile  (right panel of Fig.~\ref{fig:PV_diag}) is well below 50 AU. However, we know that the simulated disc is actually much larger, i.e. the radius of the disc with Keplerian rotation is $\sim$ 100 AU (see Fig.~\ref{fig:simulation}, bottom right panel). Therefore the simulated disc is about four times more extended than the observed one. The reason for this discrepancy is the decreasing thermal energy, i.e. the significantly lower emission, with disc radius. Hence, the true size of a protostellar disc (or at least the part of the disc which is seen in Keplerian rotation) can be significantly underestimated or -- in case a too coarse resolution is used -- no disc is detected at all. This could explain the discrepancy between observations \citep[e.g.]{Maury10,Tobin12,Hara13,Murillo13,Sanchez13,Yen13,Codella14,Maret14,Ohashi14} concerning the existence of (Keplerian) discs around Class 0 protostellar objects: Given a typical resolution of a few 0.1'' in these observations and the fact that we partly find emission coming from a region less than 50 AU in size (or 0.33'' at 150 pc; see Figs.~\ref{fig:intensity} and~\ref{fig:intensity2}), non-detections could indeed be due to a lack of resolution. As shown in Section~\ref{sec:conditions}, a resolution of 0.1'' or better may be required to properly investigate the presence of Keplerian discs around low-mass protostellar objects. Alternatively, we show that a large inclination of the disc with respect to the line-of-sight might also complicate the identification of Keplerian rotation signatures. Nevertheless, our results demonstrate that in general ALMA should be able to detect discs and to probe their dynamics under a variety of (observational and physical) conditions (see Fig.~\ref{fig:overview}).

We again note that we do not include radiative feedback in our simulations. This would result in an increased disc temperature and thus possibly stronger emission. However, since we reach temperatures up to a few 100 K in the discs even without feedback processes, we do not expect significant changes in the overall appearance of our synthetic emission maps. Furthermore, radiative feedback might alter the chemical composition of the disc, e.g. by desorption of molecules from grain surfaces. Probing this, however, would be beyond the scope of this work.

\subsection{Accuracy of protostellar masses -- fitting the ``upper'' edge}
\label{sec:upperedge}

In Section~\ref{sec:pvdiag} we have introduced a method to identify the ``upper'' edge of a Keplerian profile in a PV diagram. We show that with this method with ALMA it should be possible to determine the mass of the central object with reasonable accuracy from a PV diagram (see Tables~\ref{tab:fits} and~\ref{tab:fitsincl}). This suggests that observationally obtained protostellar masses should be relatively reliable.
Overall, for edge-on discs protostellar masses might be slightly underestimated, while they might be overestimated for inclined discs. The latter is caused by confusion of disc emission and accreting or infalling gas in the foreground.

\begin{figure}
 \includegraphics[width=\linewidth]{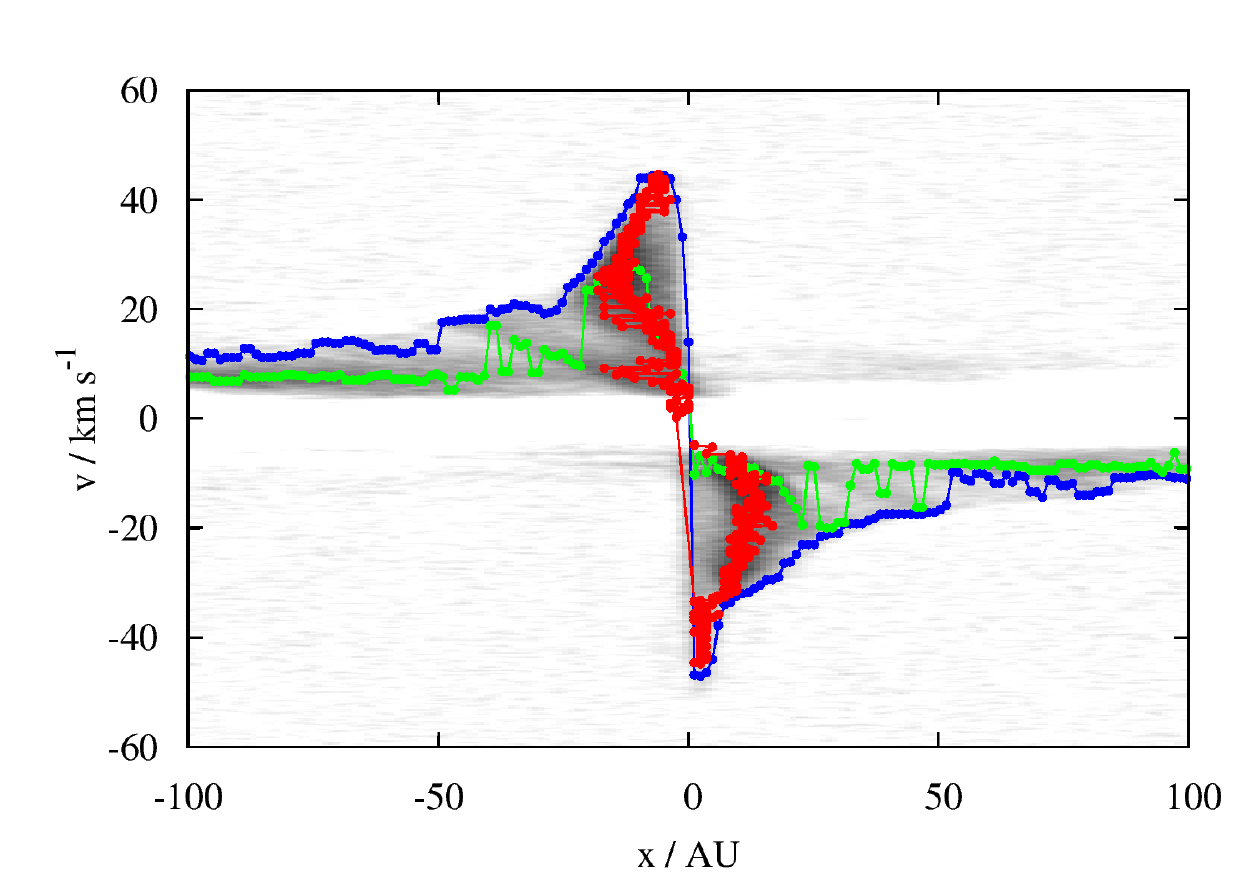}
 \caption{Comparison of different methods to determine the protostellar mass. Fitting the ``upper'' edge as done in this work (blue curve) gives the most accurate estimate. Identifying the pixel with the maximum intensity for each radius (green curve) or for each velocity channel (red curve) tends to underestimate the central mass.}
 \label{fig:upper}
\end{figure}

As explained in detail in Section~\ref{sec:upper}, our method to determine the maximum rotation velocity $v_\rmn{rot,max}$ from a PV diagram aims to fit the ``upper'' edge of the emission (see red curves in Fig.~\ref{fig:PV_diag}). A different approach widely used in literature to determine $v_\rmn{rot,max}$ is to use the pixel with the maximum intensity for each radius or alternatively for each velocity channel\footnote{More precisely a Gaussian is fitted to the intensity distribution of each velocity channel, or alternatively a Gaussian is fitted to the spectrum at each radius.} \citep{Jorgensen09,Tobin12,Hara13,Murillo13,Yen13,Codella14,Ohashi14}. Inspecting the PV diagram for $^{13}$CO $J$ = 2 -- 1 of model 1 shown in Fig.~\ref{fig:upper} shows that the pixels with the maximum intensities (red and green curve) tend to lie \textit{below} the curve identified with our method (blue curve), which in general would result in lower values for $v_\rmn{rot,max}$.
Hence, since the protostellar masses, we have obtained by fitting the ``upper'' edge, agree remarkably well with the actual mass (see Section~\ref{sec:pvdiag}), we thus expect the other methods to \textit{underestimate} the actual protostellar mass. We therefore suggest that, in order to determine protostellar masses, one should -- if possible -- try to identify the ``upper'' edge of the PV diagram rather than the pixel with the maximum intensity.

We note that the outcomes of our fitting routine are naturally somewhat dependent on the chosen threshold, which we here set to 5 times the noise level in the PV diagrams. Decreasing the threshold too drastically (e.g. to 3 times the noise level), we find that this results in a large number of outliers (pixels with high velocities), which artificially increase the protostellar mass. On the other hand, using a too high limit would define an ``upper'' edge at too low velocity values, thus leading to too small masses. Changing the threshold to 4 and 6 times the noise level results in protostellar masses which differ from the value obtained with the original limit by about 0.5 -- 2 M$_{\sun}$ for model 1 and a few times 0.1 M$_{\sun}$ for model 2. Since this uncertainty is larger than the fitting uncertainty, we therefore strongly recommend to double-check the fit by eye. However, despite this constraint, we are confident that our approach gives more accurate results than that by determining the maximum intensity. We emphasise that it has already been successfully applied to real ALMA data\footnote{F. Alves, private communication}.

\subsection{Comparison with other work}

Synthetic ALMA observations, though not using CASA, were also produced by \citet{Krumholz07} using different tracers like NH$_3$ and CH$_3$CN. In general, the emission coming from their disc appears to be somewhat more extended (a few 100 AU, see their Fig. 2) than the rather concentrated emission in this work.  However, as shown in Fig.~\ref{fig:intensity_wide}, considering only the emission coming from channels which are off by 5 -- 15 km s$^{-1}$ from the systemic velocity, the emission gets more widespread and anisotropic, thus in better agreement with the results of \citet{Krumholz07}. Moreover, we speculate that the differences in the disc sizes might be due to the fact that \citet{Krumholz07} did not include magnetic fields in their simulation, which might remove angular momentum from the centre of the molecular cloud core thus decreasing the final disc size.

More recently, in a complementary work, \citet{Dunham14} have evaluated disc masses from synthetic continuum emission maps showing that (sub-)millimetre observations usually tend to underestimate disc masses. In contrast, when taking into account the large-scale emission \citet{Harsono15} find that disc masses inferred from millimetre observations agree rather well with the actual mass. They also investigate the dynamics of a Keplerian disc by means of synthetic CO line emission maps. The authors, however, do not consider any telescope effects like incomplete coverage in the uv-plane or noise, required for mimicking ALMA observations. Nonetheless, the detection of clear Keplerian rotation profiles in their PV diagrams strongly supports our result that ALMA is able to identify the dynamics of protostellar discs in the Class 0 stage.

\subsection{Impact of chemical abundances and gas temperature}

\subsubsection{Different chemical abundances}
\label{sec:abund}

Modelling the chemical evolution of protoplanetary discs, \citet{Ilee11} find that e.g. the abundance of H$_2$CO varies between 10$^{-12}$ -- 10$^{-6}$ depending on whether gas in- or outside the disc spiral arms is considered. Moreover, they find that HCO$^+$ has an abundance of about 10$^{-11}$ in agreement with more recent simulations \citep{Evans15} and observations \citep{Jorgensen13}. Hence, using fixed fractional abundances for the different molecules (see Section~\ref{sec:synobs}) is most likely an oversimplification. We therefore consider synthetic ALMA simulations of the lines p-H$_2$CO (218.2 GHz) and HCO$^+$, $J$ = 3 -- 2 of model 1 and 2 for the edge-on case using a reduced fractional abundance of 2 $\times$ 10$^{-10}$ for  H$_2$CO and 1 $\times$ 10$^{-11}$ for HCO$^+$, i.e. lower values than used before.

From the intensity maps (see Fig.~\ref{fig:lowabundance} in the Appendix) we find that even for the lower abundances the discs are clearly recognisable. In particular for model 1 the emission is now more widespread and somewhat more noisy. Overall, for the four models considered, the total flux is decreased by roughly a factor of 3 -- 4.

Despite the lower intensities, the PV diagrams still reveal clear signs of a Keplerian rotation profile. Applying the upper-edge fits, we find protostellar masses of $12.1 \pm 0.4$ M$_{\sun}$ (HCO$^+$, $J$ = 3 -- 2, model 1), $15.4 \pm 0.7$ M$_{\sun}$ (p-H$_2$CO (218.2 GHz), model 1), $0.50 \pm 0.03$ M$_{\sun}$ (HCO$^+$, $J$ = 3 -- 2, model 2), and $0.43 \pm 0.03$ M$_{\sun}$ (p-H$_2$CO (218.2 GHz), model 2). Except maybe for the line HCO$^+$, $J$ = 3 -- 2 in model 1, these are still in reasonable agreement with the actual protostellar masses of 15.3 M$_{\sun}$ and 0.62 M$_{\sun}$.

\subsubsection{Impact of the gas temperature}
\label{sec:temp}

Since we do not  include any stellar feedback in our simulations, we might, in particular for model 1, underestimate the gas temperatures in the vicinity of the protostars \citep[e.g.][]{Krumholz07}. In order to account for this, we first calculate the luminosities of the protostars found in our simulations. For this we use the protostellar evolution model of \citet{Offner09}, which takes into account the internal stellar luminosity as well as the accretion luminosity. In a second step we then calculate the dust temperature via a Monte-Carlo simulation performed with RADMC-3D. In a final step we update the gas temperature, applying the following approach: Since the gas density is relatively high in the regions of interest (typically above 10$^8$ particles/cm$^{-3}$), it can be assumed that gas and dust are thermally tightly coupled and that their temperatures should be roughly equal. For this reason, in each grid cell we set the gas temperature to the dust temperature in case that the latter is higher. In case that the original gas temperature is higher, we leave it unchanged. In this manner, we obtain a rough estimate of the impact of stellar feedback.

We then redo the synthetic ALMA observations of the lines $^{13}$CO, $J$ = 2 -- 1 and H$^{13}$CO$^+$, $J$ = 3 -- 2 of both models for the edge-on-case. Inspecting the integrated intensity maps (see Fig.~\ref{fig:corrgastemp} in the Appendix), it can be seen that in particular for $^{13}$CO, $J$ = 2 -- 1 they overall signal-to-noise level is increased, which is also reflected by the 7 and 2 times higher total flux for model 1 and model 2, respectively. Interestingly, for H$^{13}$CO$^+$, $J$ = 3 -- 2 the obtained fluxes are somewhat lower ($\times 0.7$) for model 1 and only marginally higher ($\times 1.2$) for model 2. We attribute this to the fact that  H$^{13}$CO$^+$ is excited into higher states (e.g. the $J$ = 4 -- 3 line). However, in all cases it is possible to determine the protostellar masses from the PV diagrams, which gives masses of $17.0 \pm 0.2$ M$_{\sun}$ ($^{13}$CO, $J$ = 2 -- 1, model 1), $13.1 \pm 0.5$ M$_{\sun}$ (H$^{13}$CO$^+$, $J$ = 3 -- 2, model 1),  $0.73 \pm 0.02$ M$_{\sun}$ ($^{13}$CO, $J$ = 2 -- 1, model 2), and $0.66 \pm 0.01$ M$_{\sun}$ (H$^{13}$CO$^+$, $J$ = 3 -- 2, model 2). The values are somewhat higher than those listed in Table~\ref{tab:fits}, which we attribute to the higher signal-to-noise level in the PV diagrams. This has a similar effect as lowering the threshold value in the upper-edge fitting approach and thus leads to somewhat higher protostellar masses (see Section~\ref{sec:upperedge}).

To summarize, even though our models contain some uncertainties in the chemical composition and gas temperature, the tests performed in this section demonstrate that the general conclusion that ALMA is able to study Keplerian discs around Class 0 protostars seems to hold.

\section{Conclusions}
\label{sec:conclusions}

In this work we have presented synthetic ALMA observations of two Keplerian, protostellar discs in the Class 0 stage formed in high-resolution, 3D-MHD simulations. One of the discs contains a low-mass protostar whereas the other contains a system of 4 low- to intermediate-mass stars. We investigate how such discs appear in ALMA observations, whether the Keplerian rotation profile can be identified in PV diagrams, and what protostellar masses can be inferred. Finally, we give some guidance for observational parameters needed for future observations. The synthetic ALMA observations have been performed for the molecules $^{13}$CO, C$^{18}$O, HCO$^+$, H$^{13}$CO$^+$, N$_2$H$^+$, as well as ortho- and para-H$_2$CO, considering various transitions ($J$ values) for each molecule. In the following we summarize our main results, which where obtained assuming optimal observing conditions:
\begin{itemize}
 \item With ALMA it is/will be possible to detect Class 0 protostellar discs and velocity gradients in the 0th and 1st-order moment maps for all lines considered. In general the emission comes from the innermost part ($\leq$ 50 AU) and is thus somewhat more concentrated than the Keplerian rotation part of the discs ($\geq$ 70 AU).  We find that a resolution of a few 0.1'' might be too low to detect Keplerian discs around Class 0 objects for some low-mass protostellar systems.
 \item More extended regions can be probed by transitions with lower $J$ values, although here the signal-to-noise ratio is somewhat smaller. Transitions with higher $J$ values tend to probe emission from warmer gas closer to the protostar.
 \item Omitting the emission from channels close to the systemic velocity the accretion flow surrounding the discs can be probed. For the intermediate-mass model 1 the accretion show clearly anisotropic structures in agreement with recent observations.
 \item For most of the lines, the PV diagrams of edge-on discs reveal a clear Keplerian profile extending up to 100 AU reproducing the results of the numerical simulations. Protostellar masses determined by fitting a Keplerian rotation curve to the PV diagram are in general accurate within a few 10 percent. For the edge-on seen discs, in most of the cases the protostellar mass is slightly underestimated.
 \item 
 We show that Keplerian rotation profiles are recognisable even for strongly inclined discs (up to 30$^\circ$). Protostellar masses inferred from the PV diagrams might be overestimated in case that infall motions in the disc itself and in the fore- or background are interpreted as rotational motions. In general, however, even for strongly inclined discs the mass estimates are relatively robust. 
 \item We show that for discs seen face-on, ALMA should be able to detect spiral arms and signatures of fragmentation
\end{itemize}

We emphasise that, in order to reliably estimate protostellar masses, it is necessary to identify the ``upper'' edge of the emission in a PV diagram. In contrast, identifying the intensity maxima, a method widely used in literature, might underestimate the actual protostellar mass.
We here demonstrate the applicability of our approach and point out that it has already successfully been applied to recent ALMA observations\footnote{F. Alves, private communication}.

Finally, we summarize our results about the observational parameters considered, which can serve as a guide for future ALMA observations:
\begin{itemize}
 \item In our standard synthetic observations we assume excellent conditions, i.e. a resolution of 0.02'' -- 0.05'' (assuming a source distance of 150 pc), an observing time of 5 h, and very good weather conditions corresponding to a precipitable water vapour of 0.5 mm. Under these conditions Keplerian discs and protostellar masses can identified with high fidelity.
 \item We show that even under less optimal conditions Keplerian discs can be identified: for observing times of 2 h, moderate weather conditions (a precipitable water vapour of 1 mm), and a resolution of 0.02'' -- 0.05'', Keplerian rotation can be identified for moderately inclined discs (90$^{\circ}$ and 60$^{\circ}$). For strongly inclined discs (30$^{\circ}$) probing their dynamics is more problematic, in particular for discs around low-mass protostars.
 \item We demonstrate that even for a resolution of 0.1'' the dynamics of moderately inclined discs can be investigated, whereas for strongly inclined discs (30$^{\circ}$) Keplerian rotation cannot be probed around low-mass protostars. 
 \item Under excellent conditions (5 h observing time, 0.02'' -- 0.05'' resolution, 0.5 mm precipitable water vapour), Keplerian discs around intermediate-mass protostars can be detected even at larger distances ($\sim$ 1 kpc) and strong inclination.
\end{itemize}

\section*{Acknowledgements}

The authors like to thank the referee for his comments with helped us to significantly improve the paper. We also thank F. Alves for stimulating discussions. Furthermore, DS and SW acknowledge funding by the Bonn-Cologne Graduate School as well as the Deutsche Forschungsgemeinschaft (DFG) via the Schwerpunktprogramm SPP 1573 \textit{Physics of the ISM}. DS, \'ASM, and SW also acknowledge funding by the DFG via the Sonderforschungsbereich SFB 956 \textit{Conditions and Impact of Star Formation}. RB acknowledges funding by the Deutsche Forschungsgemeinschaft via grants BA 3706/3-1 and 3706/3-2 within the SPP \textit{The interstellar medium}, via grant BA 3706/4-1, as well as via the Emmy-Noether grant BA 3706/1-1.

\section*{Appendix A}

In order to reduce the number of figures in the main text, we here show the figures belonging to the Sections~\ref{sec:abund} and~\ref{sec:temp}.
\begin{figure*}
 \includegraphics[width=\linewidth]{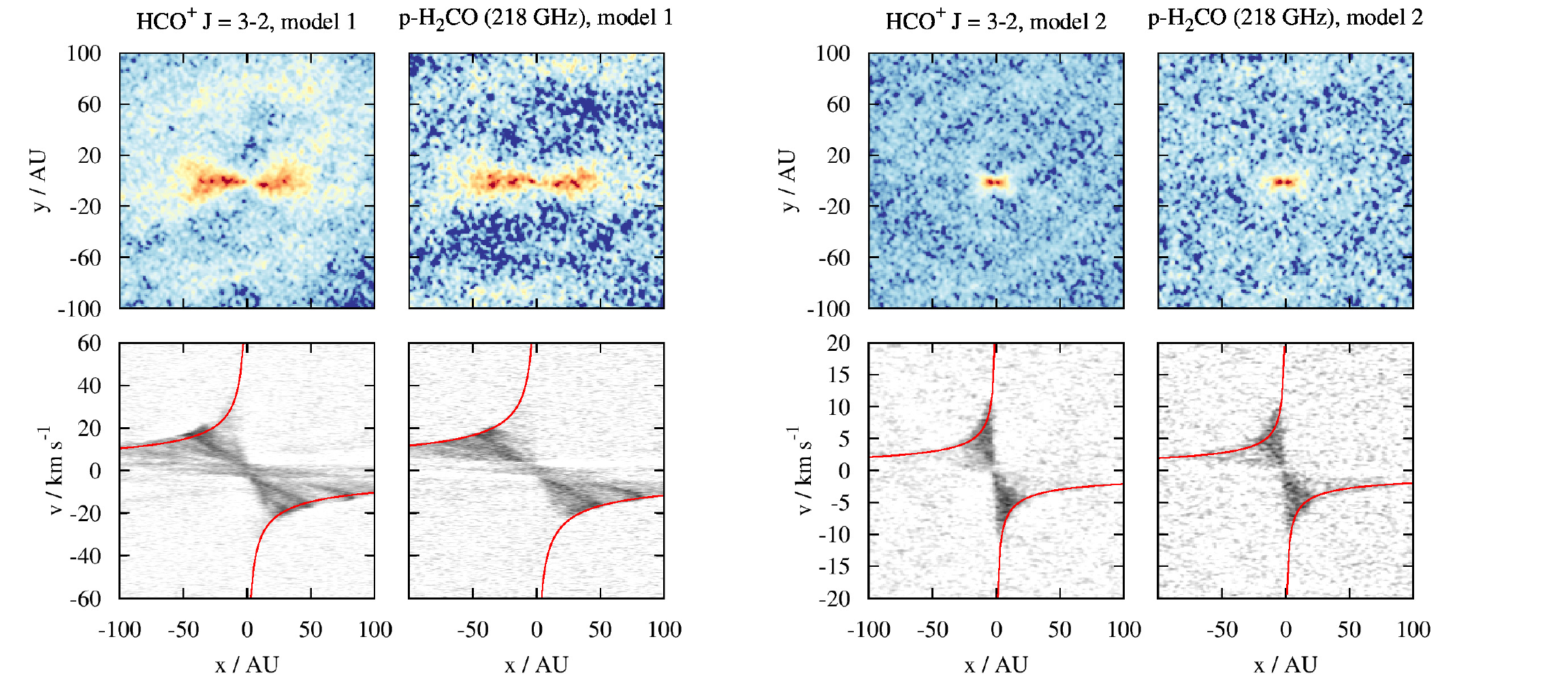}
 \caption{Intensity maps of HCO$^+$, $J$ = 3 -- 2 for model 1 and 2 under the assumption of reduced abundances (see Section~\ref{sec:abund}). The emission appears to be somewhat more widespread when comparing with Figs.~\ref{fig:intensity} and~\ref{fig:intensity2}.}
 \label{fig:lowabundance}
\end{figure*}

\begin{figure*}
 \includegraphics[width=\linewidth]{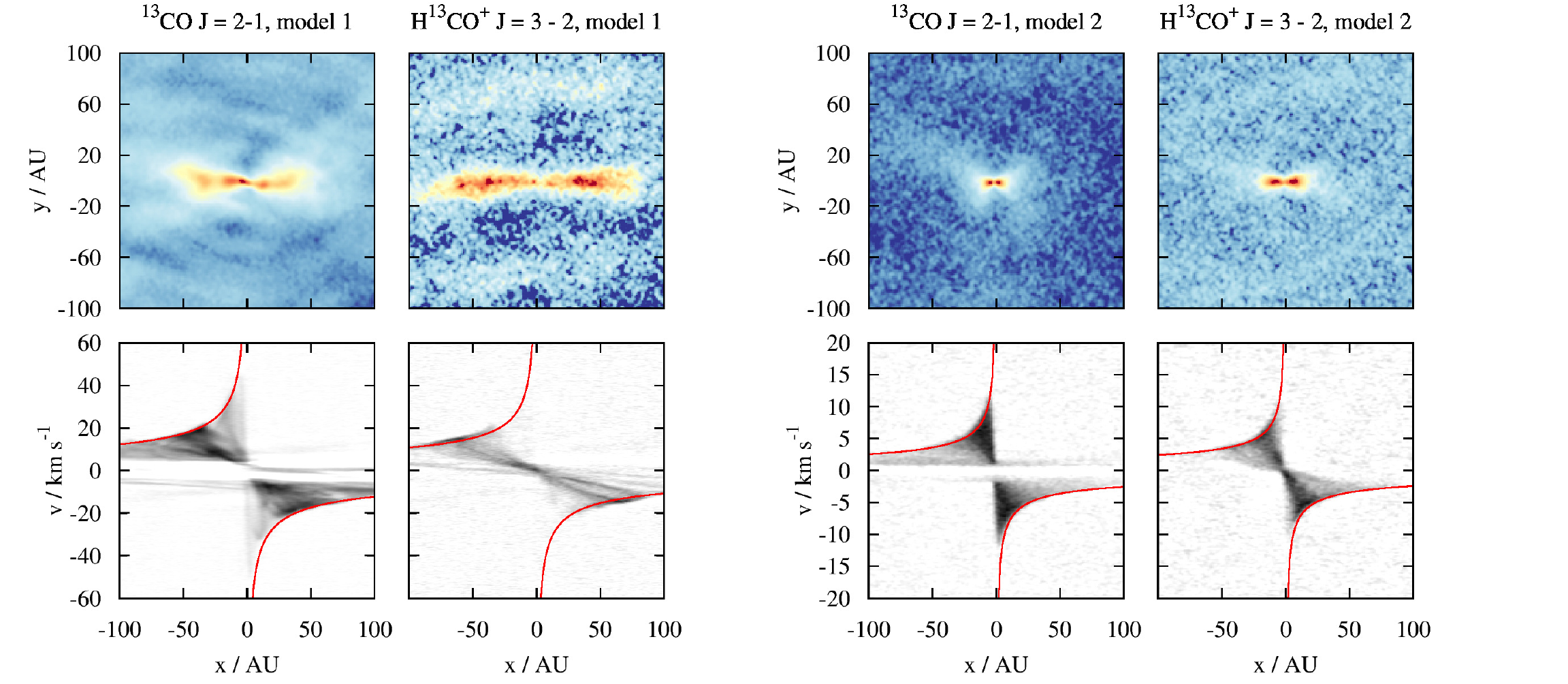}
 \caption{Intensity maps of HCO$^+$, $J$ = 3 -- 2 for model 1 and 2 under the assumption of stellar radiative feedback (see Section~\ref{sec:temp}). }
 \label{fig:corrgastemp}
\end{figure*}

\label{lastpage}

\end{document}